\documentclass[fleqn,usenatbib]{mnras}
\usepackage{newtxtext,newtxmath}
\usepackage[T1]{fontenc}
\usepackage{amsmath}
\usepackage{longtable}
\usepackage{subfloat}
\usepackage{caption}
\usepackage{subcaption}
\usepackage{mwe}
\usepackage{lscape}
\usepackage{graphicx}
\usepackage[table]{xcolor}

\usepackage{ae,aecompl}
\usepackage{pdflscape}	
\usepackage{bm}		
\usepackage{multicol}        
\usepackage{ae,aecompl}

\title[GX 5-1:AstroSat]{Anti-Correlated lags in a Neutron star Z source GX 5-1 : AstroSat's View}

\author[Chiranjeevi Pallerla \& K. Sriram]{
Chiranjeevi Pallerla,$^{1}$\thanks{E-mail: chiranjeevipallerla@gmail.com}
K. Sriram,$^{1}$
\\
$^{1}$Department of Astronomy, Osmania University , Hyderabad 500007, India\\
}

\date{Accepted XXX. Received YYY; in original form ZZZ}

\pubyear{xxxx}

\begin{document}
\label{firstpage}
\pagerange{\pageref{firstpage}--\pageref{lastpage}}
\maketitle

\begin{abstract}
We report the cross-correlation function studies of a Neutron star low mass X-ray binary, a Z source GX 5-1 using SXT and LAXPC energy bands onboard AstroSat. For the first time, we report the lag between soft (0.8-2.0 keV, SXT) and hard X-ray energy bands (10-20 keV and 16-40 keV, LAXPC) in GX 5-1 and detected lags of the order of a few tens to hundreds of seconds in the Horizontal branch. We interpreted them as the readjustment time scale of the inner region of the accretion disc. We used various two components and three-component spectral models to unfold the spectra and observed the changes in soft and hard component fluxes which were exhibiting Horizontal Branch Oscillation variations. It was observed that the bbody component assumed to be originating from the boundary layer over the NS, was also found to vary along with the HBO variation where lags were detected. We constrained the size of the Comptonizing region of the order 15-55 km assuming that lags were due to variation in the size of the corona. We noticed a similar size of the Comptonizing region after employing other models and suggest that the overall size of corona must be of the order of a few tens of km to explain the lags, HBO variation, and respective spectral variations. In a case study, it was noted that the BL size increases as GX 5-1 vary from the top of the HB to the upper vertex.
\end{abstract}

\begin{keywords}
accretion, accretion disc - stars: neutron star -- X-rays: binaries -- X-rays: individual: GX 5-1
\end{keywords}

    \section{INTRODUCTION}

Neutron star X-ray binaries are an important class of low mass X-ray binaries to understand the radiative and dynamical configuration of the inner region of an accretion disc. Though from previous studies especially based on RXTE (Rossi X-ray Timing Explorer) data of Z sources, it was known that there must exist a corona / Comptonization region to explain the observed hardtail in their X-ray spectra but the exact location and how it changes across the intensity variation is not yet properly understood. Among the two primary categories i.e. Z and Atoll sources, Z sources emit close to the Eddington luminosity (0.5—1.0 L$_{Edd}$; \citet{2007A&ARv..15....1D}) and they exhibit "Z" and "C" shape intensity variation in the Hardness Intensity diagram (HID) or colour-colour diagrams (CCDs) \citep{hasinger1989two, 2006csxs.book...39V}. The Z shape variation constitutes a Horizontal branch (HB) at the top, a Flaring branch (FB) at the bottom and a Normal branch (NB) connecting them diagonally. These are further classified into two broad groups viz. Sco and Cyg-like sources due to their different appearance exhibited by the HB and FB i.e. less vertical orientation of HB and a weaker FB is seen among Cyg-like sources than in Sco-like \citep{1994A&A...289..795K}.  The hybrid source XTE J1701--462 occupies a special place among NS LMXBs and is considered to be a remarkable source as it displays all the characteristics exhibited by both Z and atoll sources \citep{Homan_2010,homan2007rossi, Lin_2009}. At the brightest state, the intensity variations were associated with HB, NB, and FB of Cyg-like and exhibited Sco-like variation at relatively lower brightness. During the decay phase, the variation closely resembles the soft state of an atoll source and later transits to the hard state of the atoll source just before going to the quiescent state. Many important results were noticed based on the spectral fitting of RXTE data of this source. The mass accretion rate was found to be constant along with the Z phase in Sco-like variation and different mechanisms were proposed to explain the spectral and timing variations during the Z phase variations  \citep{Lin_2009}. It was also found that mass accretion rate is the important driving parameter during the Z and all along with the atoll phases variation.
Z sources are unique probes in the sense they provide a platform to understand the structure of accretion disc emitting close to Eddington luminosity because due to the radiation pressure the structure of the inner region of accretion is affected.   The previous studies suggested that the interplay between the accretion disc and Comptonization region mutually varies to produce the observed tracks in the HID. However other physical components like a boundary layer \citep{popham2001accretion} or a transition layer (TL) \citep{osherovich1999kilohertz, osherovich1999interpretation, titarchuk1999correlations} cannot be ruled out. The Comptonization region can be in the form of a quasi-spherical cloud or it could be a base of a jet that causes the observed hard continuum in the X-ray spectrum \citep{migliari2007linking}. But its association with dynamical features like various branch oscillations or band-limited noises is not known. The spectra of Z sources can also be explained by a structure known as the boundary layer over the NS surface but again, its association to the observed HBO, NBO etc. is not properly understood. \citep{popham2001accretion, gilfanov2003boundary, 2006A&A...453..253R}. Based on the detailed spectral modelling of GX 17+2, BL occupies a smaller area at the lower vertex (i.e. bottom of NB) in comparison to its area in other branches \citep{lin2012spectral} and the Comptonization dominates at the HB branch that fades away as source traverse to the FB. The inner disc radius was found to be moving towards the NS as the Z track evolves from HB to FB.  All these structural and radiative variations are found to be occurring at an almost constant mass accretion rate \citep{Lin_2009, lin2012spectral}.

One of the major issues is the size of the Comptonization region at HB and NB and how it is associated with the various spectral and dynamical properties. Across the HB, as the HBO frequency increases toward the upper vertex of NB, the Comptonization region decreases \citep{lin2012spectral, sriram2012anti} but it is somehow connected to the inner disc radius but the exact association is not known \citep{lin2012spectral}. The Cross-Correlation Function (CCF) provides a method by which one can constrain the size of the Comptonization region. It is been noticed that anti-correlated and correlated hard and soft lags of a few tens to a hundred seconds were observed in Z and atoll sources between RXTE PCA's soft and hard energy band light curves of the order \citep{sriram2012anti, sriram2019constraining}. These lags were interpreted as the readjustment time scales of corona or disc occurring in the inner region of the accretion disc.

Recently, observations based on AstroSat satellite also unveiled such lags in GX 17+2 and 4U 1705-44 using LAXPC and SXT observations \citep{malu2021investigating, malu2021exploring, malu2020coronal, sriram2021understanding}. \citet{sriram2019constraining} also reported a 113 $\pm$ 51 s anti-correlated hard lag based on the NuSTAR data for GX 17+2. Based on both timing and spectral results it was proposed that these lags are readjustment time scales of the Comptonization region during which its size was changing \citep{malu2021investigating, sriram2021understanding}.  Interestingly for the first time, the CCF between soft (0.8-2.0 keV) of AstroSat’s SXT and hard 16-40 keV (LAXPC) light curves of GX 17+2 revealed the anti-correlated lags in the NB. Moreover using the lags, we constrained the height of the corona to the order of a few tens of km which was well matching the sizes obtained from other models and the inner disc radius was found to be close to the last stable orbit. \citet{sriram2012anti} reported lags in GX 5-1 and those were mostly associated with the HB and NB where the Comptonization region is needed to explain both timing and spectral behaviour.

The X-ray continuum of Z sources can be modelled by soft and hard components. The soft component is arising from the Keplerian portion of the accretion disc \citep{mitsuda1984energy} or originates from the surface of the NS or a boundary layer \citep{lin2012spectral}. The hard component is probably arising from a scattering of the soft photons in the Comptonization region located in the inner region of the accretion \citep{agrawal2009x, sriram2012anti, di2001probing, done2007modelling}. Most often the structure of the Comptonization region is assumed to be quasi-spherical but this is not always the case. Sometimes an extended corona is needed to explain X-ray spectra and the ionized emission line in these sources \citep{2014MNRAS.438.2784C, schulz2009heating}. The other structure that can potentially explain the hard component is the presence of a boundary layer (BL) on the surface of NS associated with 7-20 keV in the X-ray continuum \citep{miller2013constraints, ludlam2017truncation}. Sometimes a hybrid model is often invoked to explain the spectra across the various branches along the Z track and the relative modulation of parameters is used to understand the variation in the physical structures \citep{lin2007evaluating, lin2012spectral}.

GX 5-1 is the second brightest persistent LMXB source in the direction of the galactic center \citep{bradt1968celestial} and its X-ray variations are similar to a Cyg-like source with an extended HB. It emits close to the Eddington luminosity i.e. L/L$_{Edd}$ =1.6-2.3 \citep{jackson2009model, homan2018absence}. It exhibits HBO, NBO and Kilo Hertz QPOs across the HB and NB with no QPO features in FB. \citep{van1985intensity, lewin1992quasi, wijnands1998millisecond}. Based on the simultaneous coupled appearance of HBO and disappearance of NBO, a possible puffed-up inner disc model was proposed \citep{sriram2011coupled}. Based on the spectral studies, \citet{homan2018absence} did not find any signature of reflection feature in the spectra observed from the NuSTAR satellite, probably due to the presence highly ionized disc. Radio observations revealed the presence of jet in GX 5-1 \citep{penninx1989non, tan1992simultaneous, fender2000radio} which is often detected in other Z sources eg. GX 17+2 \citep{migliari2007linking}.

In this paper, for the first time, we report the anti-correlated lags between soft and hard energy bands obtained from SXT and LAXPC onboard AstroSat. We also report the spectral variations and changes in the HBO features observed during detected lag time scales. Finally, we estimate the size of the Comptonization region at HB using various models and also constrain the inner disc radius in GX 5-1.

    \section{DATA REDUCTION AND ANALYSIS}
        
We used AstroSat's Large Area X-ray Proportional Counter (LAXPC) and Soft X-ray telescope (SXT) archival data of the source GX 5-1. The Observation ID of the Data set was G06-114T01-9000001056 starting from 2017, February 26 to 2017, February 27 (23690 s). The LAXPC has three identical proportional counter units LAXPC 10, LAXPC 20, LAXPC 30 spanning a total effective area of $\sim$ 6000 cm$^2$ at 15 keV operating in the energy range of 3-80 keV  \citep{yadav2016large, agrawal2017large, antia2017calibration}. The total effective area of SXT is $\sim$ 128 cm$^2$ at 1.5 keV and operates in the 0.3-8.0 keV energy range \citep{singh2017soft}. To study the power density spectrum (PDS), we used the LAXPC data in the Event Analysis (EA) mode with a timing resolution of 10 $\mu$s. SXT has an energy resolution of 150 eV and Photon Counting (PC) mode has a time resolution of 2.4 s.

We reduced LAXPC level 1 data to level 2 data using the LAXPC software (Format A, Aug 04, 2020 version) \footnote{\url{http://Astrosat-ssc.iucaa.in/?q=laxpcData}}. This software is provided by AstroSat Support Center (ASSC). For reducing level 1 data to level 2 data we have followed the steps using LAXPC software $\tt LAXPC\_MAKE\_EVENT$ generating the event file and $\tt LAXPC\_MAKE\_STDGTI$ generating the good time interval. GTI files (Earth occultation and SAA removed), event files, and filter files were produced. For producing light curves and spectra, the following programs were used viz. $\tt LAXPC\_MAKE\_LIGHTCURVE$ and $\tt LAXPC\_MAKE\_SPECTRA$.

SXT level 2 data was used for timing and spectral analysis. A merged cleaned event file was generated by event merger code and an ancillary response file was generated by sxtARFmodule. These are made available on the SXT website. The image, light curves, and spectra were generated by XSELECT V2.4k. SXT team has provided the response file sxt\_pc\_mat\_g0to12.rmf and sky background spectrum file SkyBkg\_comb\_EL3p5\_Cl\_Rd16p0\_v01.pha which were used during the spectral reduction \footnote{\url{https://www.tifr.res.in/~astrosat_sxt/dataanalysis.html}}. The source was extracted with an aperture size of 5.5$^\prime$ -- 12.0$^\prime$ annulus region to avoid the pile-up effects\footnote{\url{https://www.iucaa.in/~astrosat/AstroSat_handbook.eps}}.

    \section{TIMING ANALYSIS}
    
The Figure 1 show the LAXPC 3.0-20.0 keV (top) and SXT 0.3-8.0 keV light curves. The p1 to p6 markers in figure 1 show the locations in the HID (Fig. 2.)  We extracted the hardness intensity diagram (HID, Figure 2) using the LAXPC 10 data where the hard colour is defined as 8.0-20.0 keV / 3.0-8.0 keV and intensity was calculated in the energy range of 3.0-20.0 keV. Boxes show the various p1 to p6 where the lags were detected as the source traverses from NB to HB branches. For the first time, we report the cross-correlation functions (CCF) of GX 5-1 between soft, 0.8-2.0 keV, and hard, 10-20 keV, and 16-40 keV energy bands of SXT and LAXPC instruments onboard the AstroSat. For the CCF analysis, LAXPC light curves were divided into the following sections i.e. sections A to P, Table 1 and figure 3. Various sections and their meaning in terms of energy bands and duration are shown in Table 1. For example the light curve belonging to p1 region in HID was split into sections A $\&$ B to perform cross-correlation functions (CCFs) using SXT (0.8-2.0 keV) and LAXPC (10-20 keV and 16-40 keV) energy bands (see Table 1). The soft X-ray light curves were almost steady when compared to the variability of the corresponding hard X-ray light curves.  The CCF was performed using the program crosscor made available in XRONOS package (for more details see \citealt{sriram2007anticorrelated, sriram2021understanding}). We selected the simultaneous observed light curve segments observed from SXT and LAXPC.  We used LAXPC 10 data with a bin time of 32 s to perform the CCFs and LAXPC 10, LAXPC 20, and LAXPC 30 for power density spectrum analysis. It is evident that CCFs are complex and hence to constrain the lag, we fitted the portion of the CCF where the anti-correlation coefficients are high. A Gaussian function was fitted to the lag portion of the CCF and corresponding errors were estimated with a criterion $\Delta \chi^2$ =2.71 at 90\% confidence level. Each of the figure (Figure 3) display the soft and hard light curves (left: top and bottom panels) and the corresponding CCFs were shown in the right panels along with the fits at the lag in the inset of each figure. The shaded region (red) is the standard deviation of the CCF function and the vertical line marks the zero lag in each figure. All the correlation coefficient and lag values along with their error bars are reported in Table 1. In section A, an anti-correlated soft lag of -197 $\pm$ 17 s is been observed between 0.8-2.0 keV and 10-20 keV energy bands whereas a lag of -230 $\pm$ 27 s was found with a slightly higher energy band (16-40 keV) but within error bars. In sections A and D, the CCFs are between SXT and LAPXC (10-20 keV) energy bands, and sections B, C, and E are the CCFs between SXT and LAPXC (16-40 keV) energy bands.

\begin{figure}
\centering
\caption{Top: The background subtracted SXT (0.3-8.0 keV) and LAXPC (3.0 - 20.0 keV) light curves are shown. The bin size is 2.3778 s for both SXT and LAXPC. The arrow markers P1 to P6 represents the location of the LAXPC light curve in the HID.}
\includegraphics[width=7cm, height=9cm, angle=270]{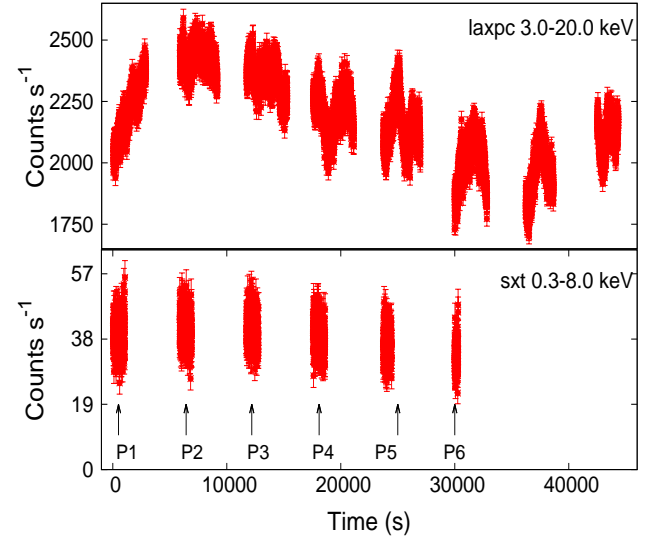}
\end{figure}

\begin{figure}
\centering
\caption{The Hardness Intensity Diagram  (HID) of GX 5-1 was observed with the AstroSat's LAXPC 10 instrument. Hard colour is defined as 8.0-20.0 / 3.0-8.0 keV and intensity were calculated with a bin size of 32 s in the 3.0-20.0 keV energy range. The horizontal and normal branches shows can be depicted in the figure. The boxes p1 to p6 show the source variation during the observation where lags were detected. Segment A belongs to region p4, segment B and segment C belong to region p5, and segment D belongs to p5 and p6 regions in the HID.}
\includegraphics[width=7cm, height=9cm, angle=270]{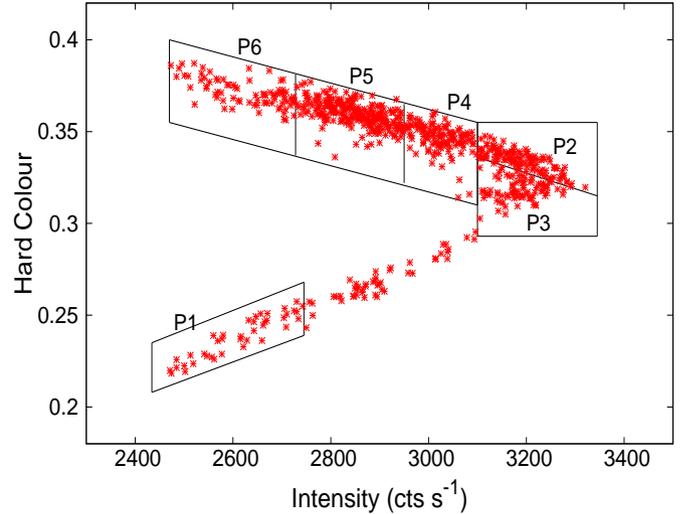}
\end{figure}

From sections F to P, the CCFs are performed between 3-5 keV and 16-40 keV (LAXPC), shown in Figure 3. In section F, the CCF positively peaked at zero lag. For other sections, we found anti-correlated hard and soft lags of about a few tens to hundreds of seconds except for section N's CCF where a correlated hard lag was noticed (see Table 1). From our previous studies \citep{sriram2012anti, sriram2019constraining, malu2021investigating, sriram2021understanding} we argue that the observed lags are the readjustment time scale during which the inner region of the accretion disc, where the Comptonization structure or inner disc radius/ boundary layer is possibly changing in its size. 

If this is the case then one should also observe variations in the HBO's centroid frequency during the detected lags. \citet{bhulla2019astrosat} reported HBOs around 35 Hz along with NBO around 7 Hz in this source using AstroSat data.  We searched for such variations by studying the power density spectrum (PDS) in each of the segments. We detect variation in HBO frequencies of four light curves of LAXPC. The fourth \&  fifth segments of the LAXPC light curve are termed as segments A \& B and the seventh \& eighth light curves are represented as segments C and D (see figure 1). To detect maximum variation in HBO frequencies in a segment, we took the maximum and minimum count rate light curve sections and extracted the PDS in the energy range 3-20 keV with a bin time of 1/512 s.  Each PDS was fitted with a power law for the band-limited noise and a Lorentzian model for HBO. PDS was normalized in units of (rms/mean)$^2$/Hz \citep{1991ApJ...383..784M}. Figure 4 shows that light curves and vertical lines mark the maximum and minimum count rate sub-sections for which PDSs were studied (left top panel, 3-20 keV). The corresponding PDSs along with the fits are shown in the left-bottom panel in each figure along with a vertical line marking the shift of the centroid frequency of the HBO. In each figure, the right panels (top and bottom) show the corresponding spectra (discussed in more detail, see section 4). We noted that on four occasions, we noticed variations in the centroid frequency (see Table 2). The maximum variation in the frequency is been noticed for segments A and C i.e. $\sim$44 Hz to $\sim$35 Hz and $\sim$ 27 Hz to $\sim$ 35 Hz. These shifts strongly suggest that during the observed lag, the Comptonization region varied.

\begin{figure*}
\caption{The background-subtracted SXT (0.8-2.0 keV), LAXPC soft (3-5 keV) and hard (16-40 keV) light curves are shown in left panels. The right panel shows the corresponding cross-correlation function. The cross-correlation function (CCF) of each section of the light curve and shaded regions show the standard deviation of the CCFs. The inset figure gives the Gaussian fit of the lag portion of the CCF.}

\begin{tabular}{ c @{\quad} c }
\begin{subfigure}[b]{0.5\textwidth}
\includegraphics[width=8.5cm, height=5cm, angle=0]{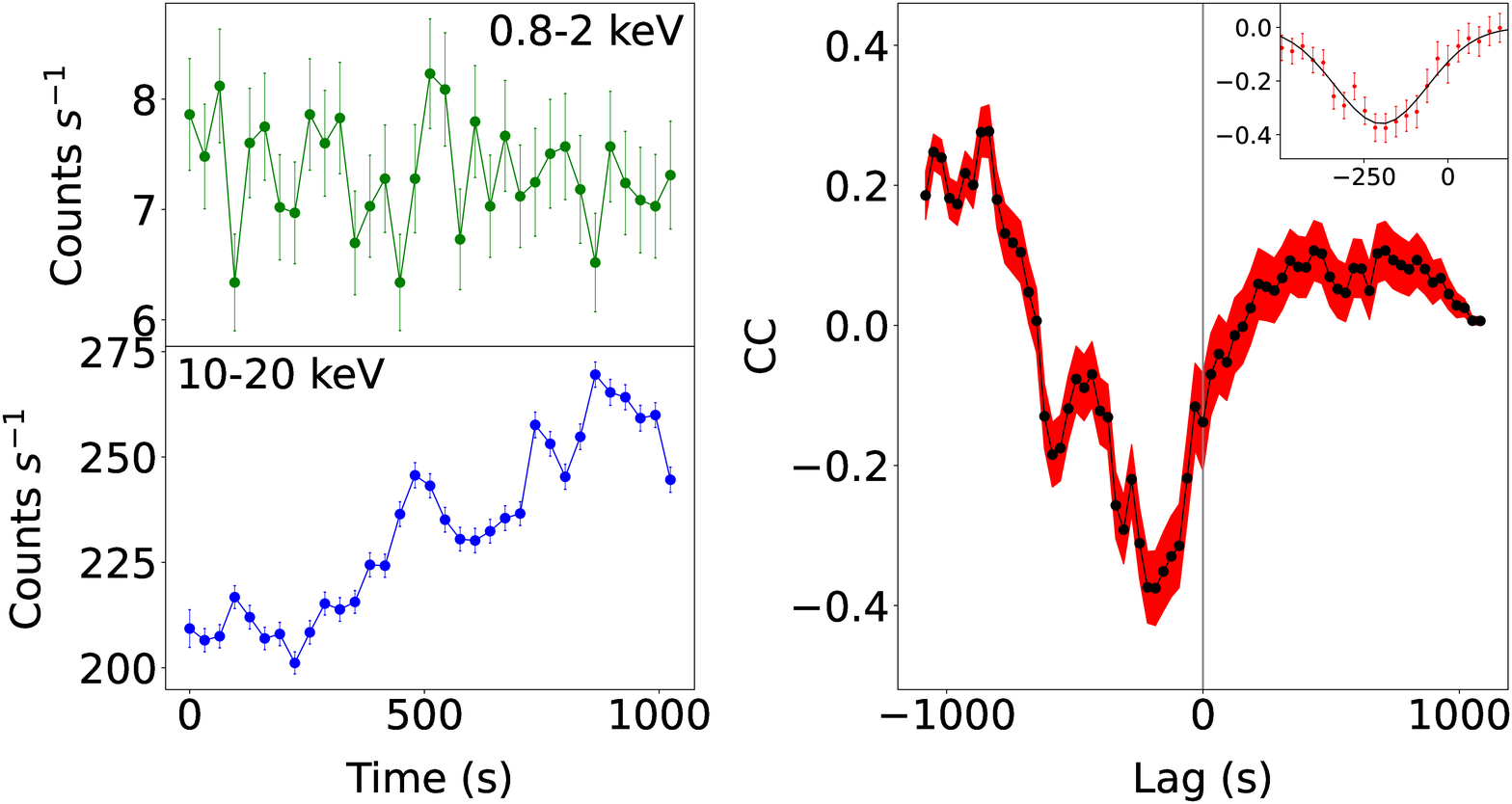}
\caption{Section A} 
\end{subfigure}

\begin{subfigure}[b]{0.5\textwidth}
\includegraphics[width=8.5cm, height=5cm, angle=0]{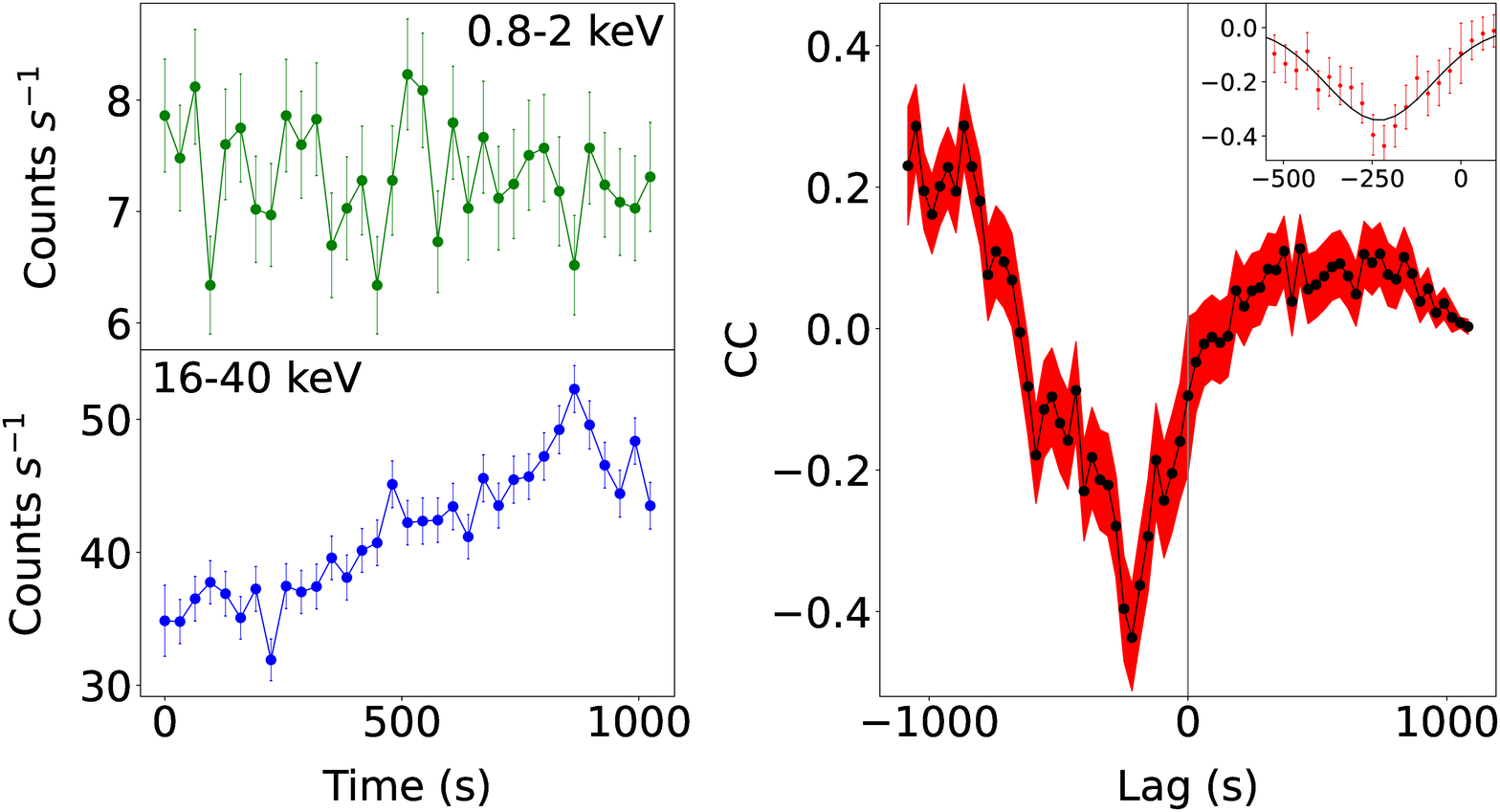}
\caption{Section B}
\end{subfigure}
\end{tabular}

\begin{tabular}{ c @{\quad} c }
\begin{subfigure}[b]{0.5\textwidth}
\includegraphics[width=8.5cm, height=5cm, angle=0]{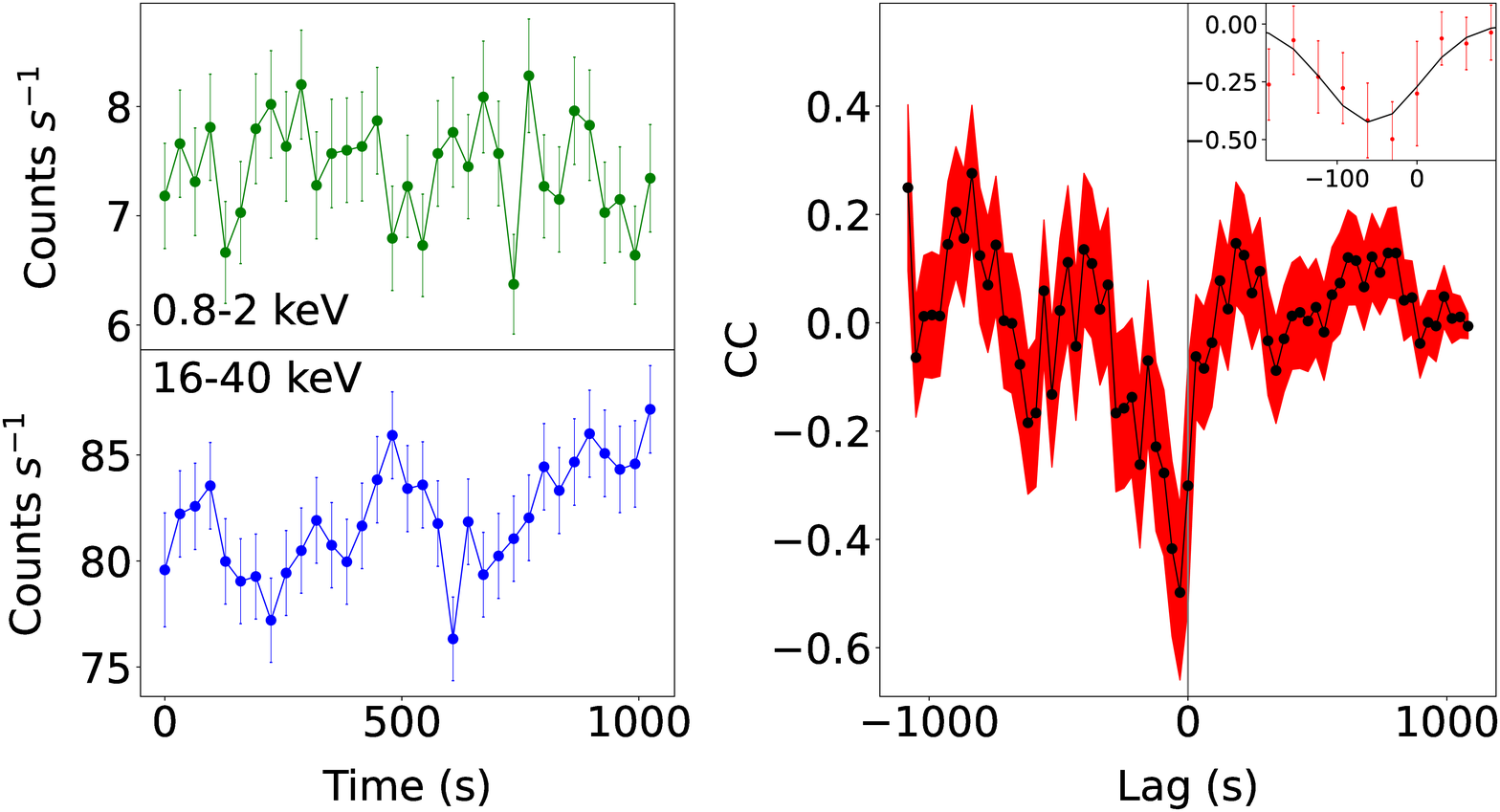}
\caption{Section C}
\end{subfigure}

\begin{subfigure}[b]{0.5\textwidth}
\includegraphics[width=8.5cm, height=5cm, angle=0]{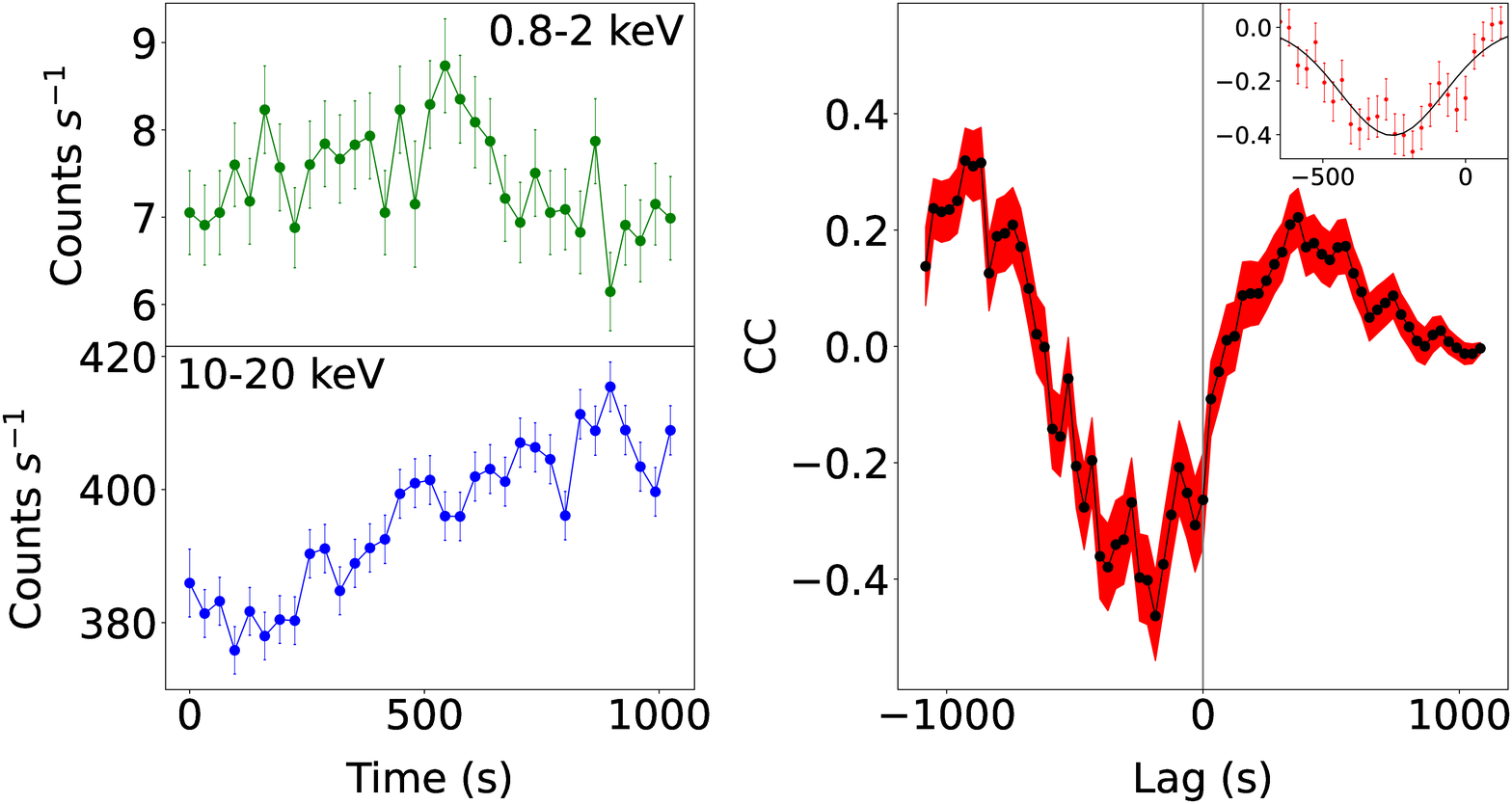}
\caption{Section D}
\end{subfigure}
\end{tabular}

\begin{tabular}{ c @{\quad} c }
\begin{subfigure}[b]{0.5\textwidth}
\includegraphics[width=8.5cm, height=5cm, angle=0]{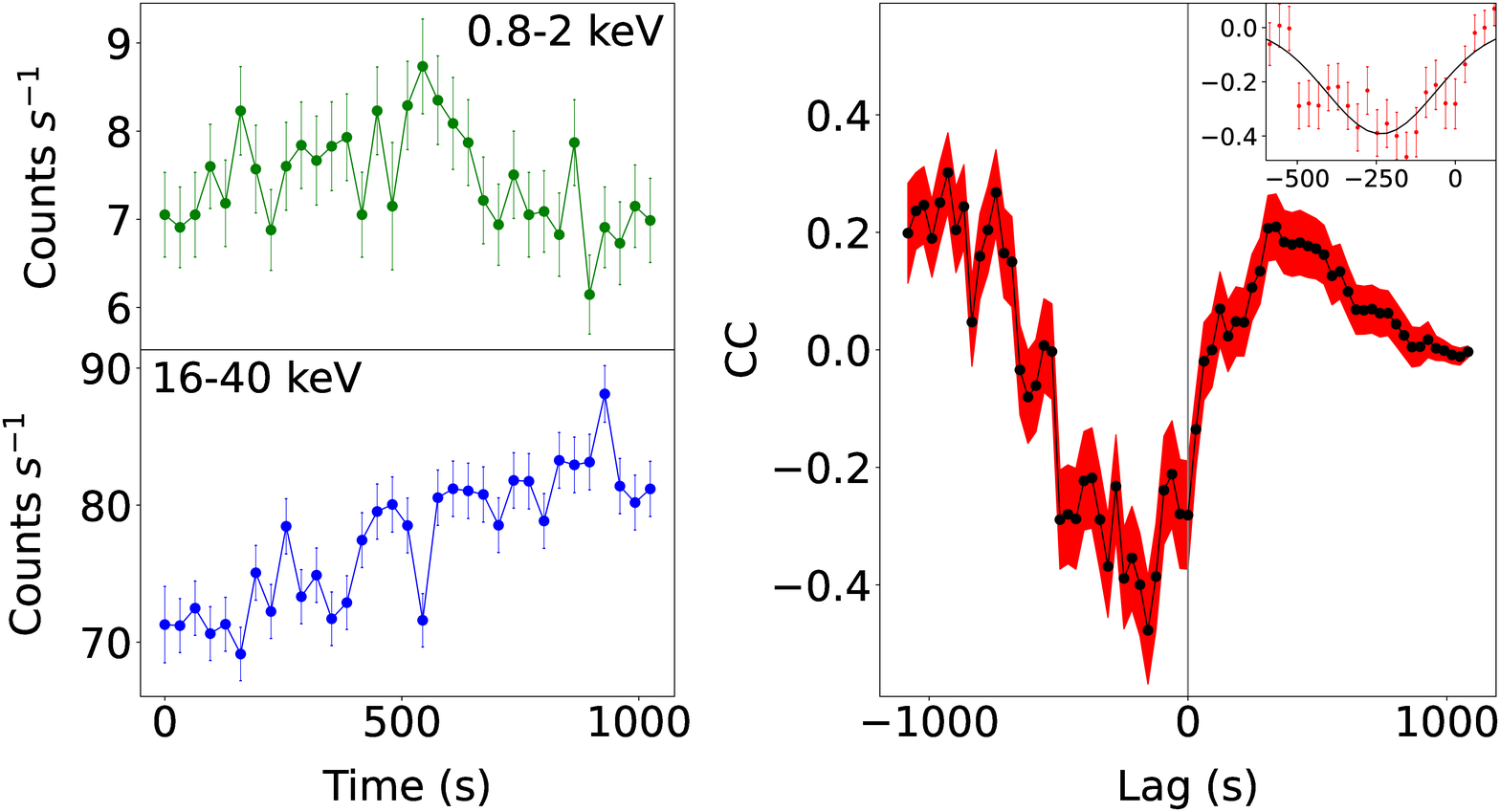}
\caption{Section E}
\end{subfigure}

\begin{subfigure}[b]{0.5\textwidth}
\includegraphics[width=8.5cm, height=5cm, angle=0]{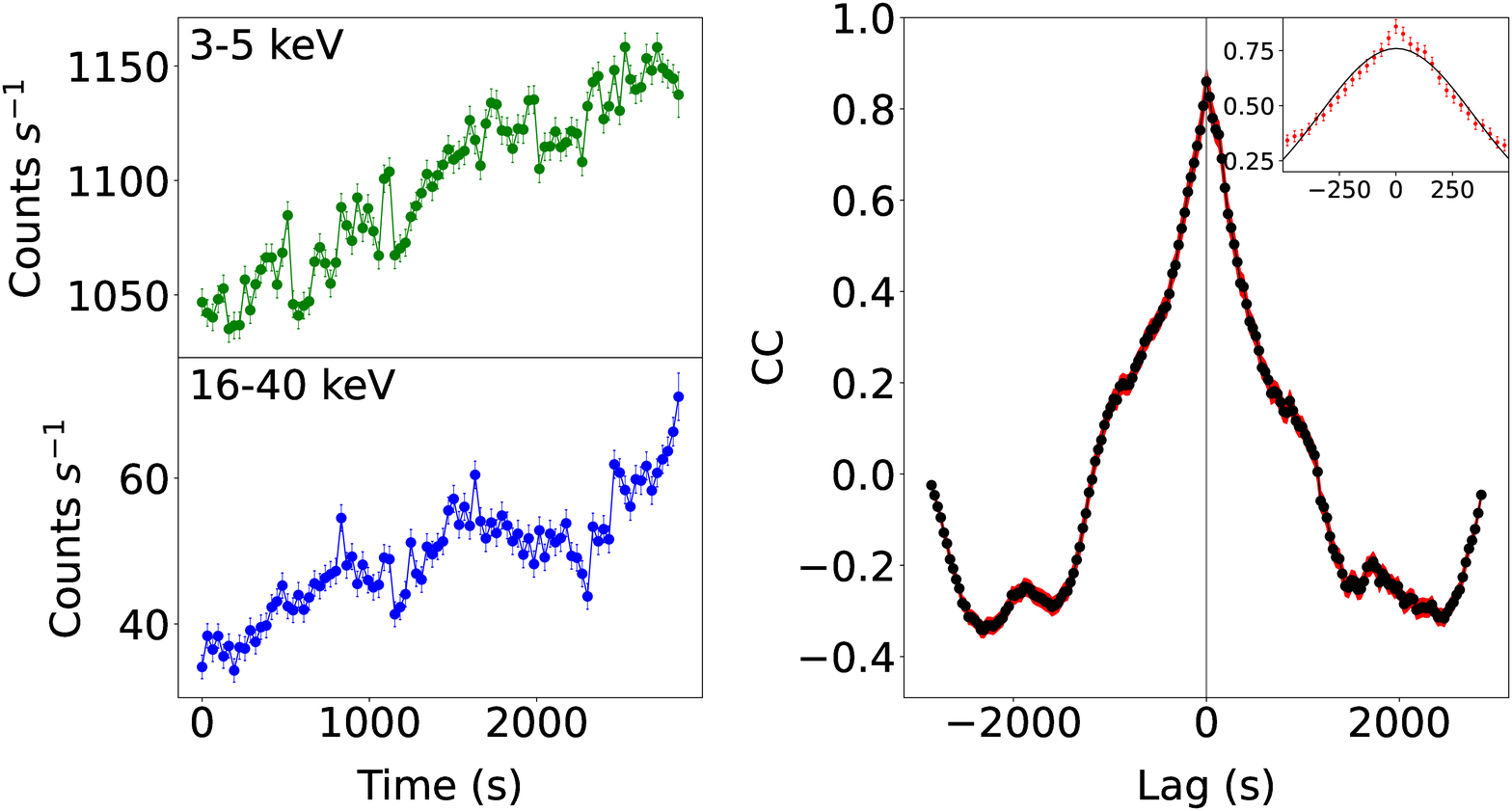}
\caption{Section F}
\end{subfigure}
\end{tabular}

\begin{tabular}{ c @{\quad} c }
\begin{subfigure}[b]{0.5\textwidth}
\includegraphics[width=8.5cm, height=5cm, angle=0]{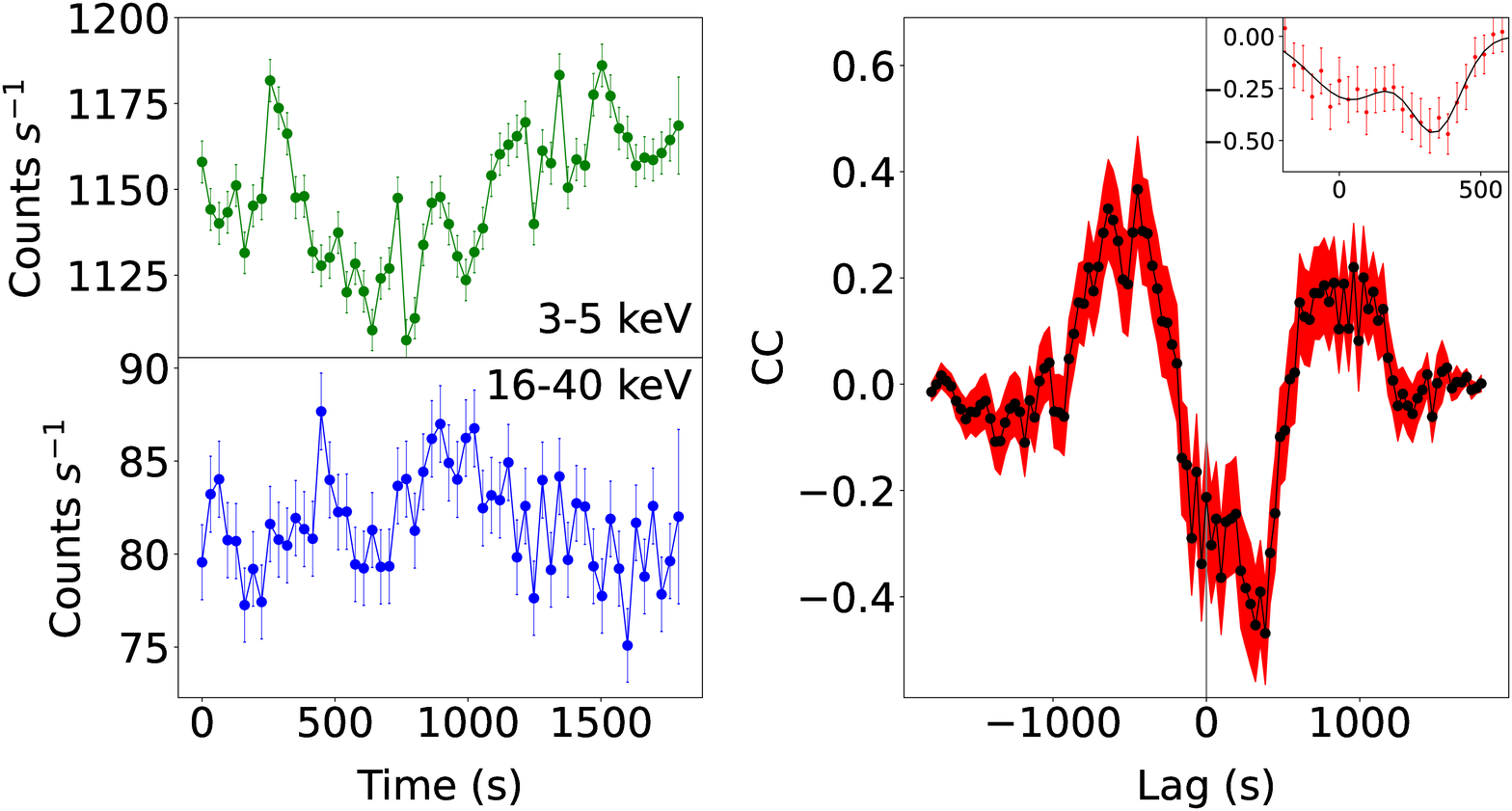}
\caption{Section G}
\end{subfigure}

\begin{subfigure}[b]{0.5\textwidth}
\includegraphics[width=8.5cm, height=5cm, angle=0]{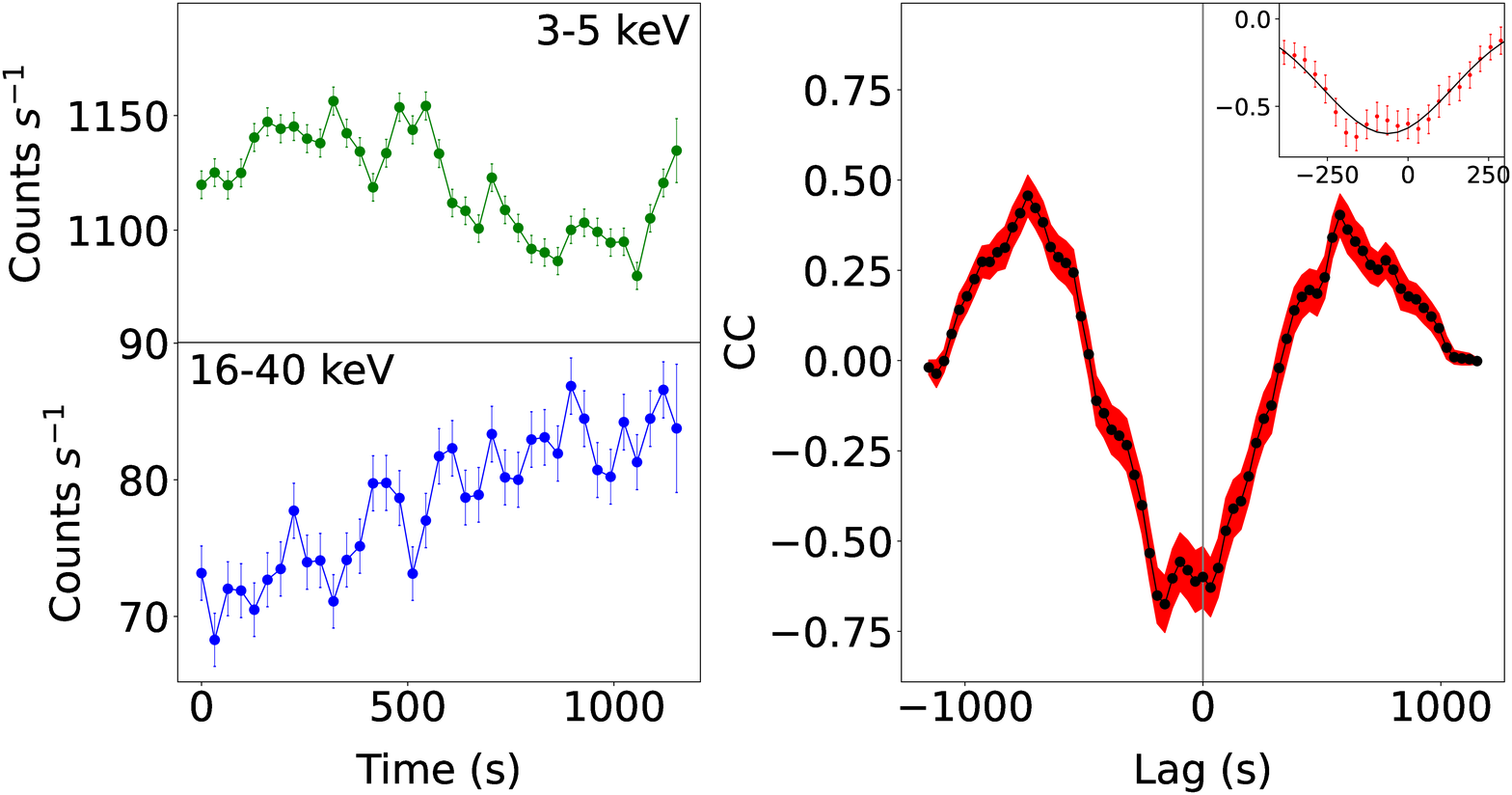}
\caption{Section H}
\end{subfigure}
\end{tabular}
\end{figure*}

\begin{figure*}\ContinuedFloat

\begin{tabular}{ c @{\quad} c }
\begin{subfigure}[b]{0.5\textwidth}
\includegraphics[width=8.5cm, height=5cm, angle=0]{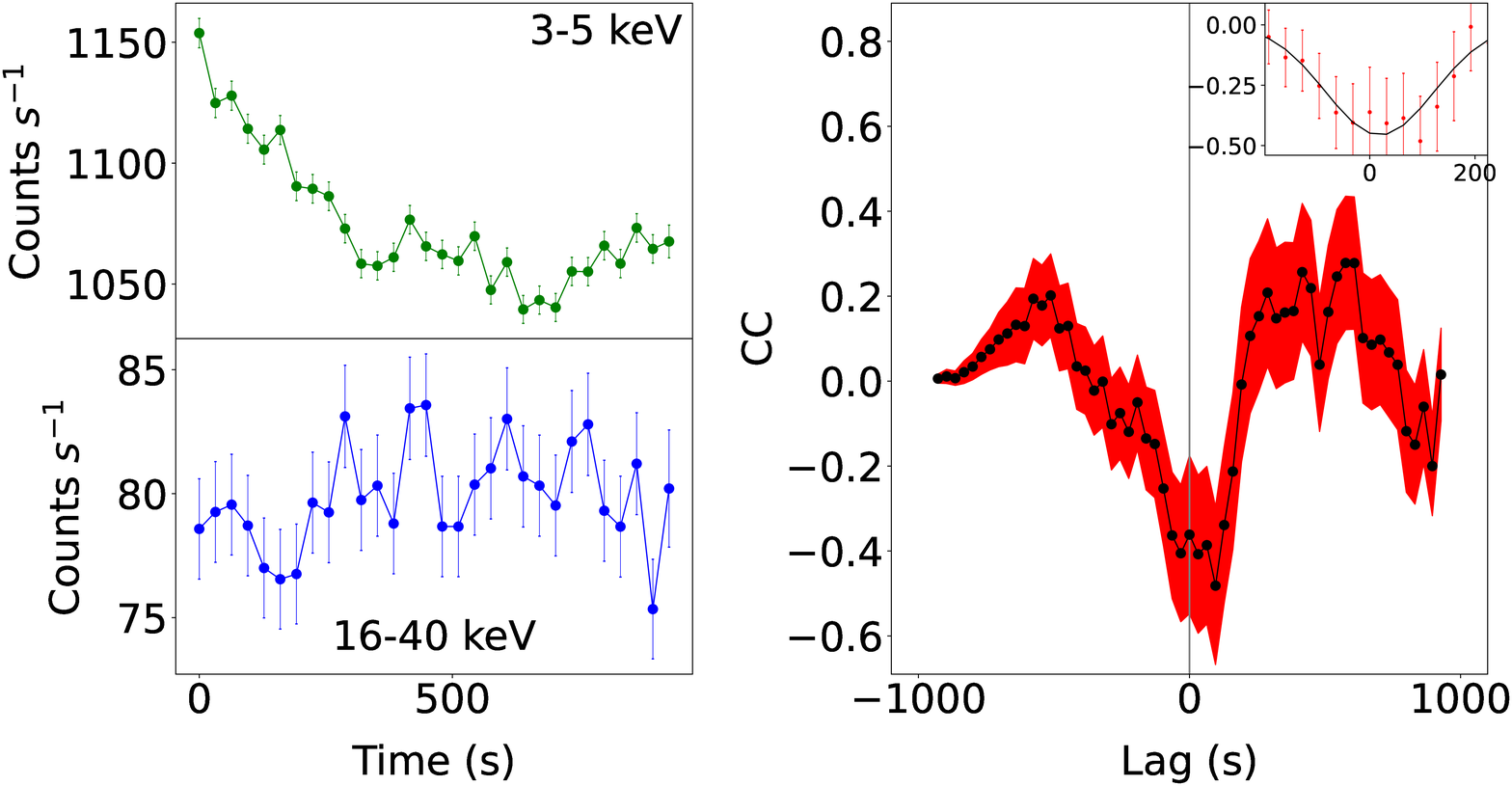}
\caption{Section I}
\end{subfigure}

\begin{subfigure}[b]{0.5\textwidth}
\includegraphics[width=8.5cm, height=5cm, angle=0]{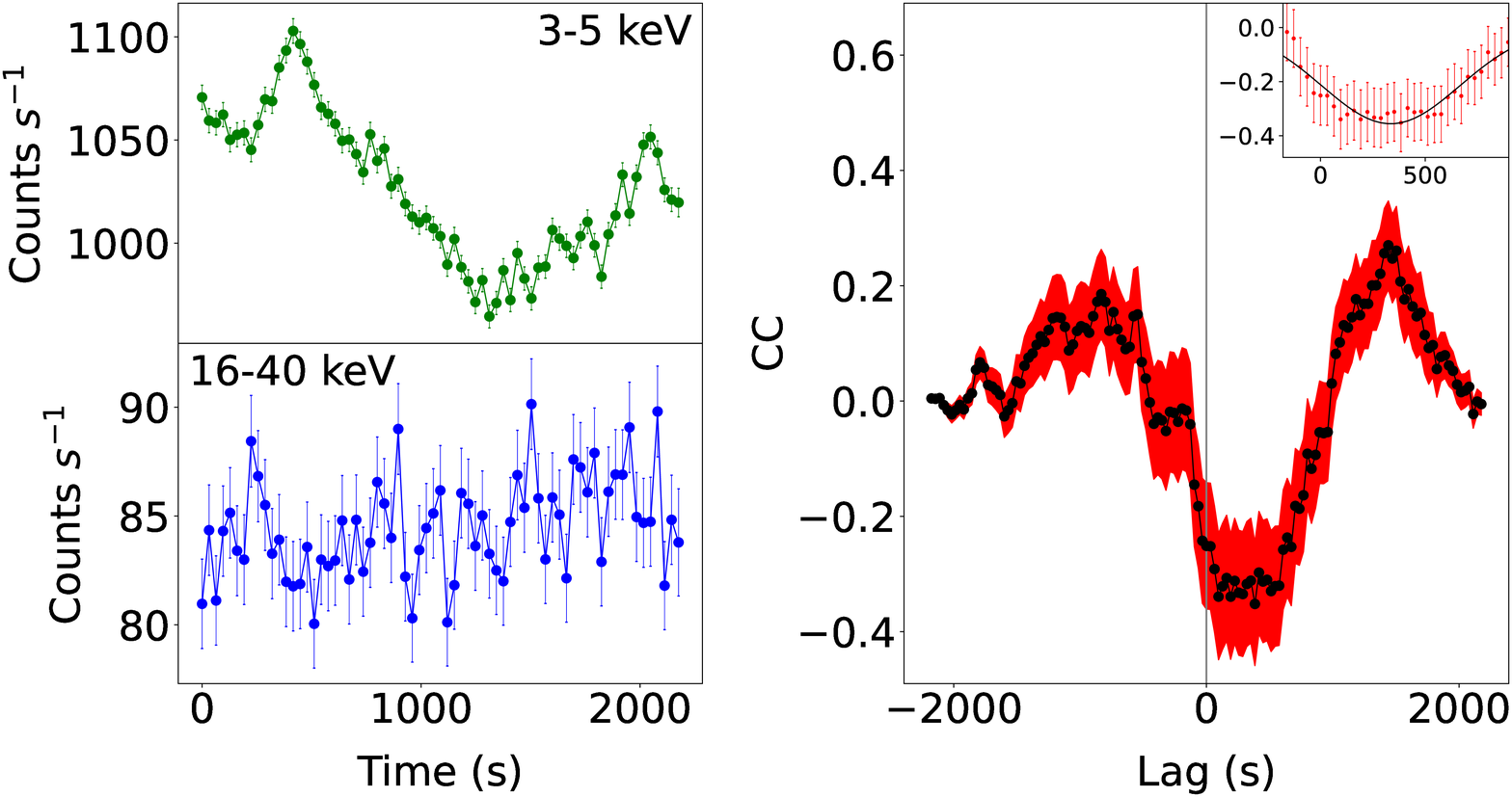}
\caption{Section J}
\end{subfigure}
\end{tabular}

\begin{tabular}{ c @{\quad} c }
\begin{subfigure}[b]{0.5\textwidth}
\includegraphics[width=8.5cm, height=5cm, angle=0]{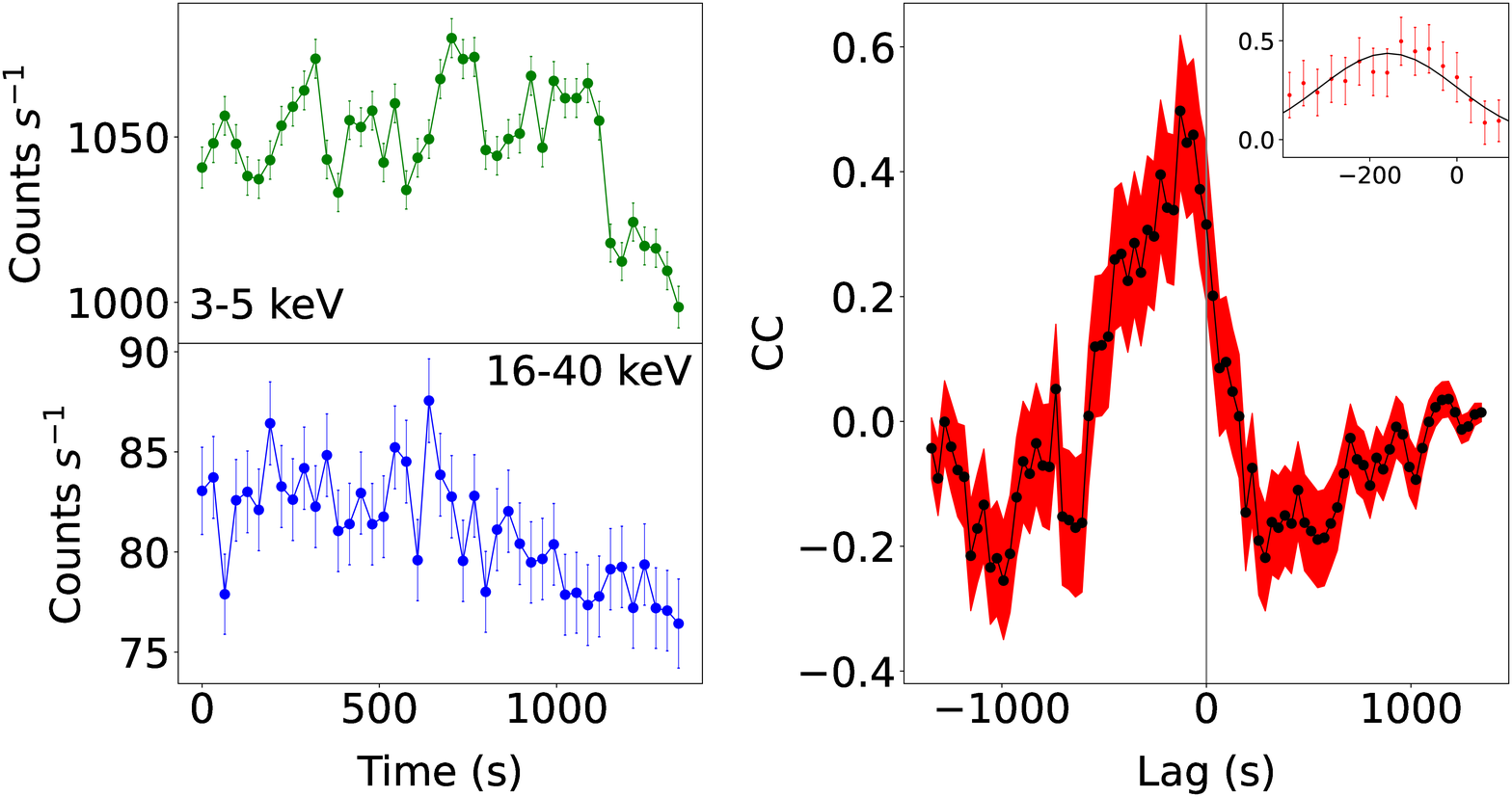}
\caption{Section K}
\end{subfigure}

\begin{subfigure}[b]{0.5\textwidth}
\includegraphics[width=8.5cm, height=5cm, angle=0]{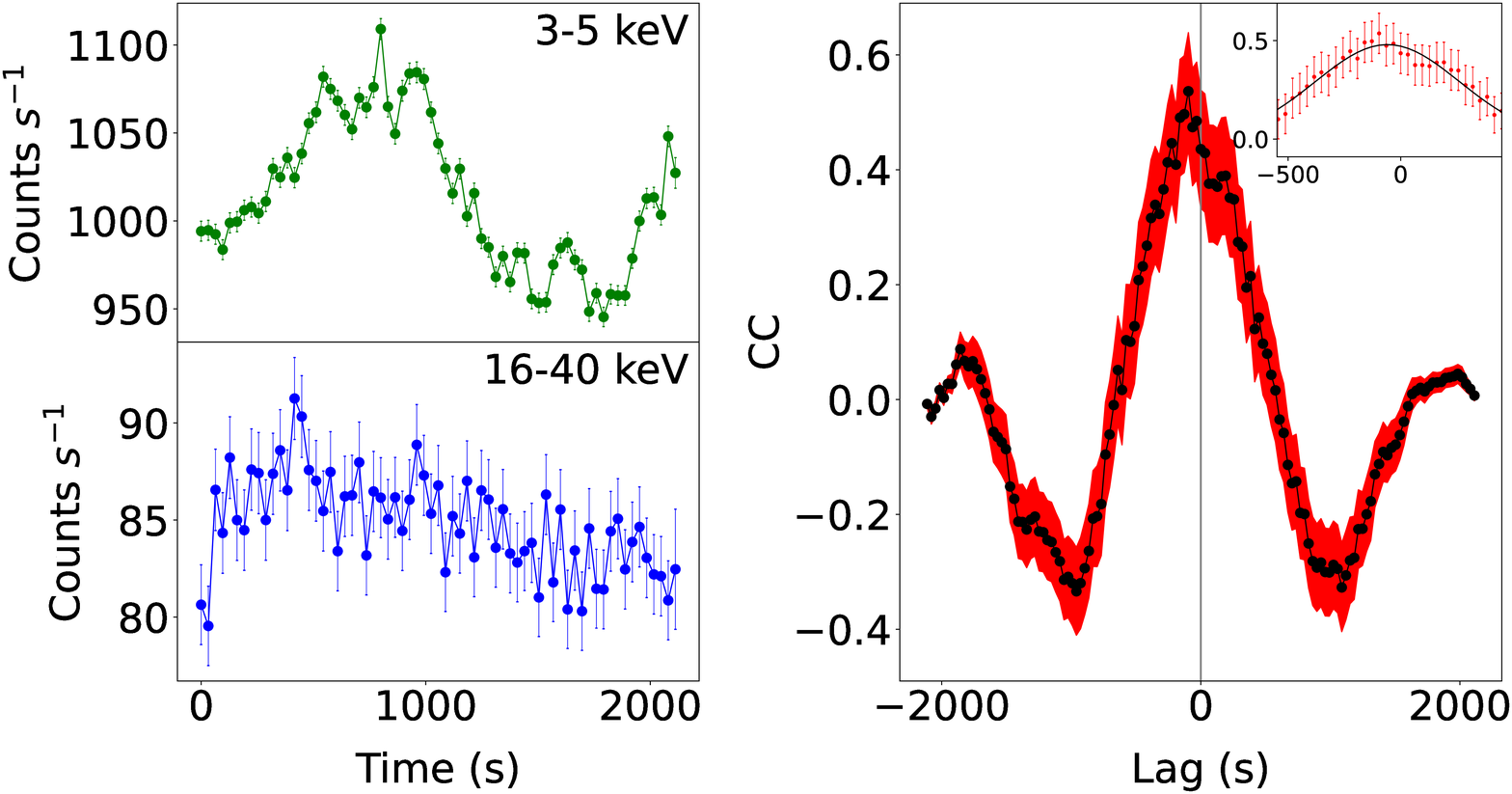}
\caption{Section L}
\end{subfigure}
\end{tabular}

\begin{tabular}{ c @{\quad} c }
\begin{subfigure}[b]{0.5\textwidth}
\includegraphics[width=8.5cm, height=5cm, angle=0]{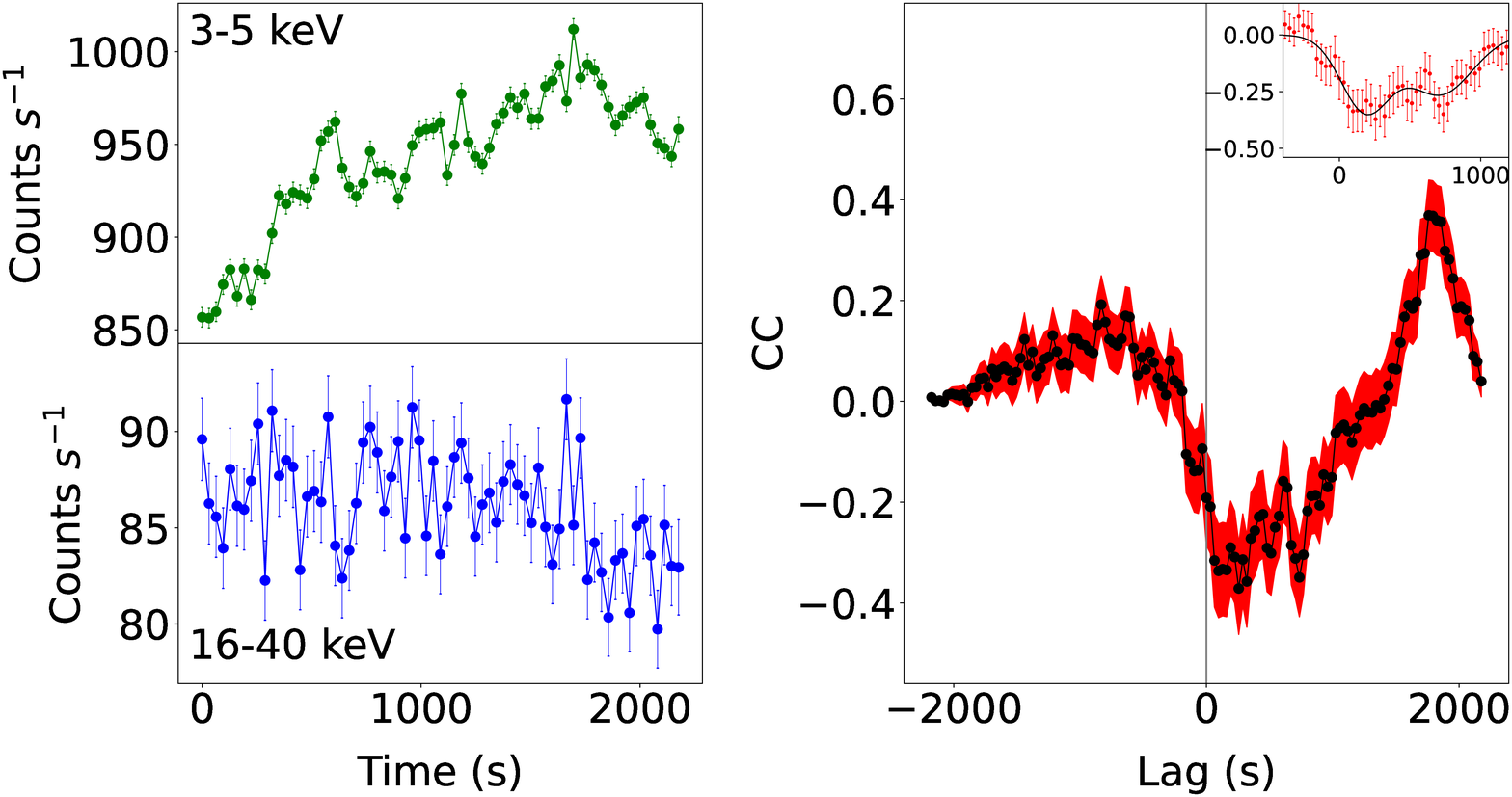}
\caption{Section M}
\end{subfigure}

\begin{subfigure}[b]{0.5\textwidth}
\includegraphics[width=8.5cm, height=5cm, angle=0]{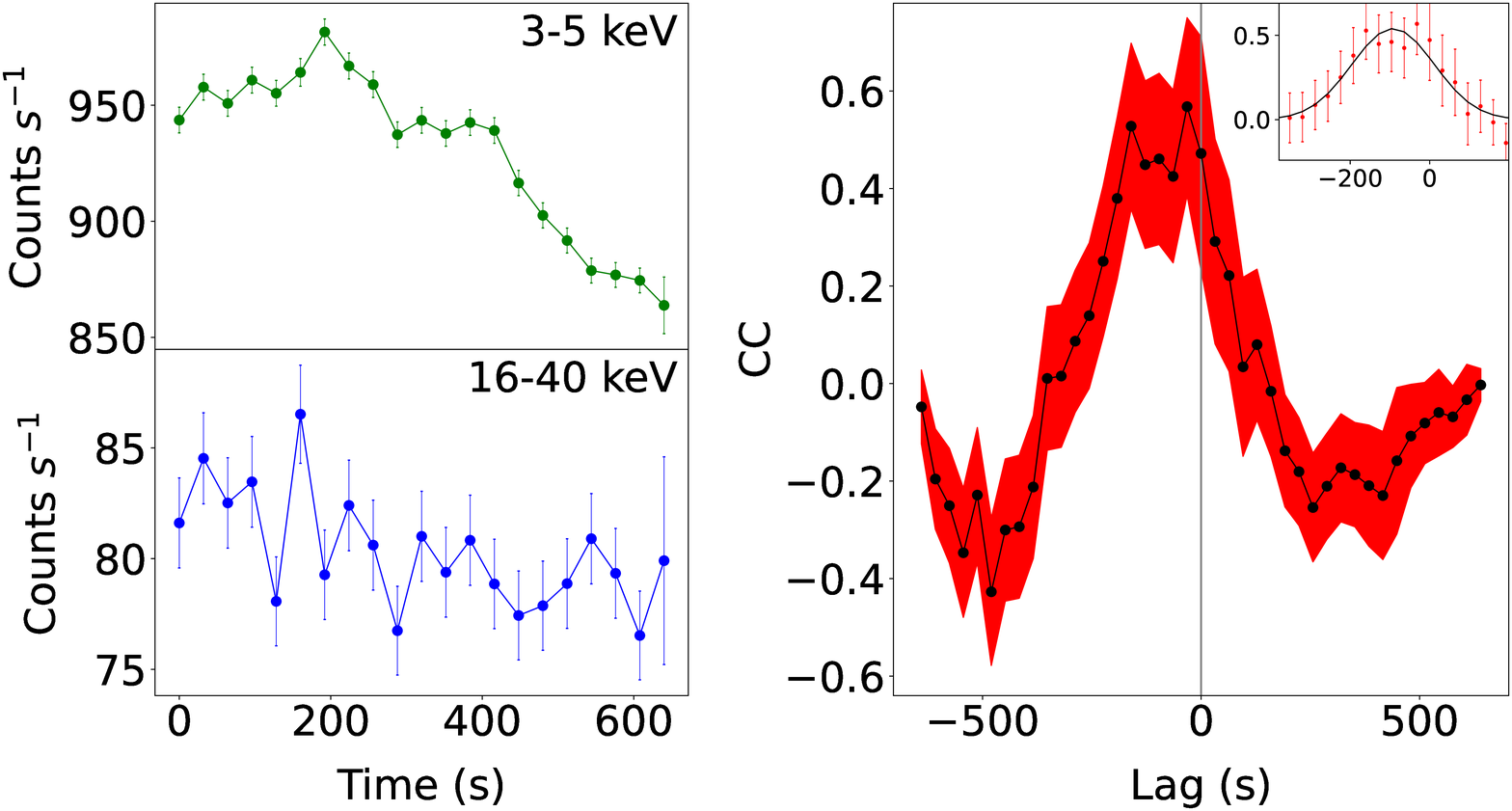}
\caption{Section N}
\end{subfigure}
\end{tabular}

\begin{tabular}{ c @{\quad} c }
\begin{subfigure}[b]{0.5\textwidth}
\includegraphics[width=8.5cm, height=5cm, angle=0]{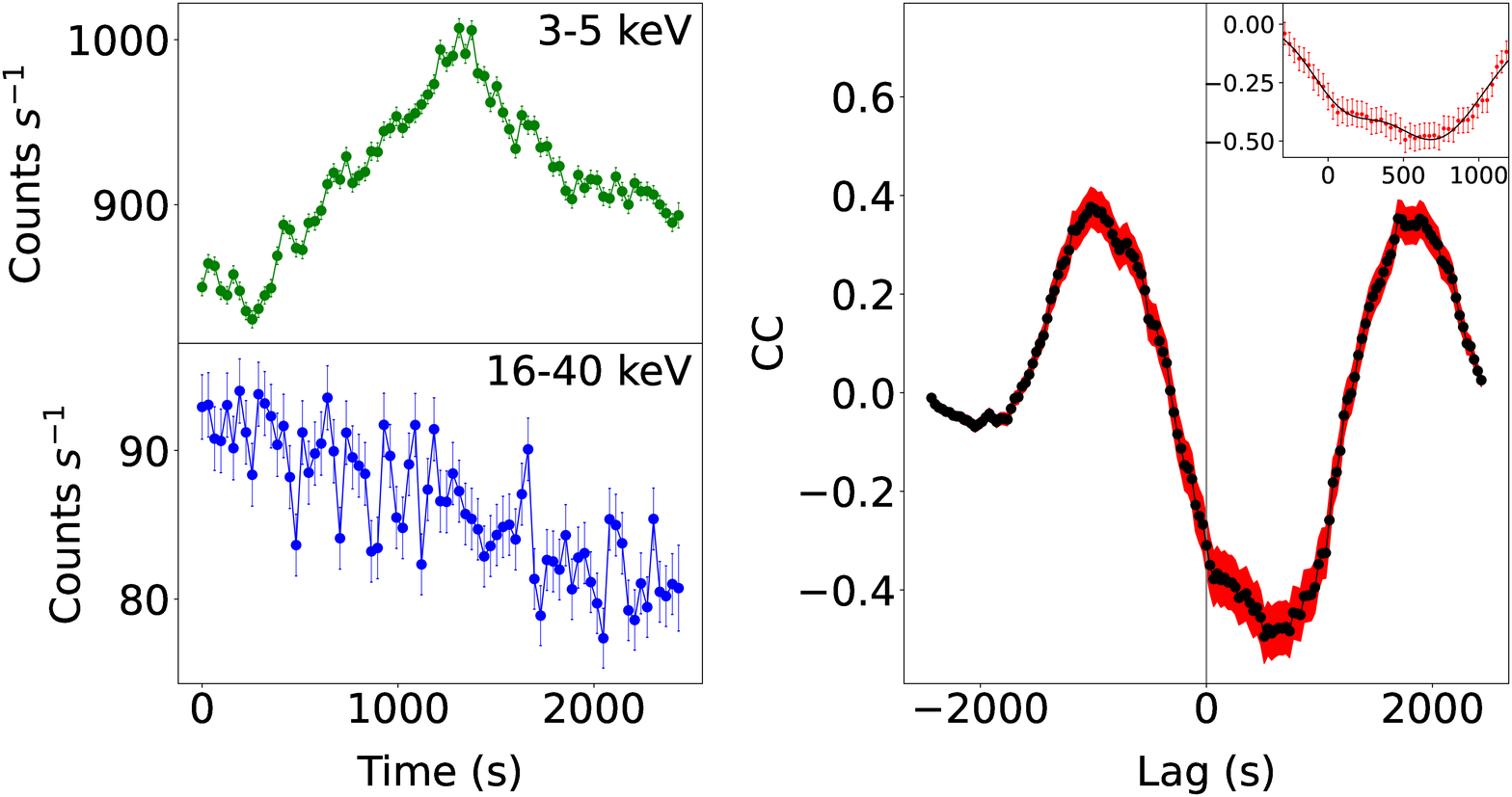}
\caption{Section O}
\end{subfigure}

\begin{subfigure}[b]{0.5\textwidth}
\includegraphics[width=8.5cm, height=5cm, angle=0]{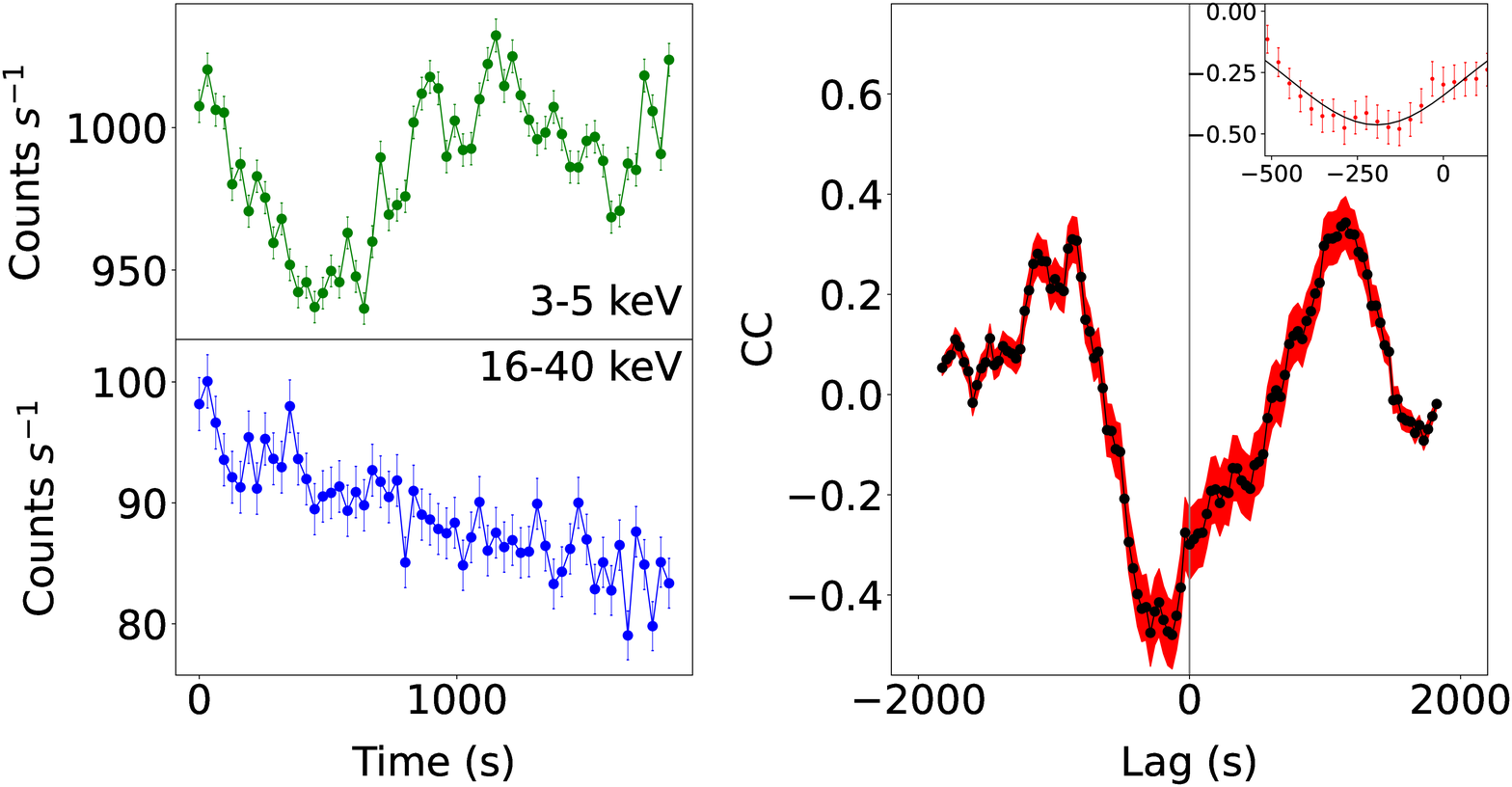}
\caption{Section P}
\end{subfigure}
\end{tabular}
\end{figure*}

\begin{figure*}
\centering
\caption {In each figure, the 3--20 keV energy band light curve is shown in the left top panel, and the power density spectra of two sections marked in the light curve are shown in the bottom panel. The unfolded spectra along with their model components (model: bbody+compTT) and corresponding residuals of fit of the two sections (HBO1 and HBO2) are shown in the right panels (for more details see text). The connection between sections and segments are  follows: section J and section K belongs to segment A, section L  belongs to segment B, section O belongs to segment C, and section P belongs to segment D. The spectra Y-axis are in the units of  keV$^2$ (Photons cm$^{-2}$ s$^{-1}$ keV$^{-1}$)}.

\begin{tabular}{ c @{\quad} c }
\begin{subfigure}[b]{0.5\textwidth}
\includegraphics[width=9.3cm, height=7.5cm, angle=0]{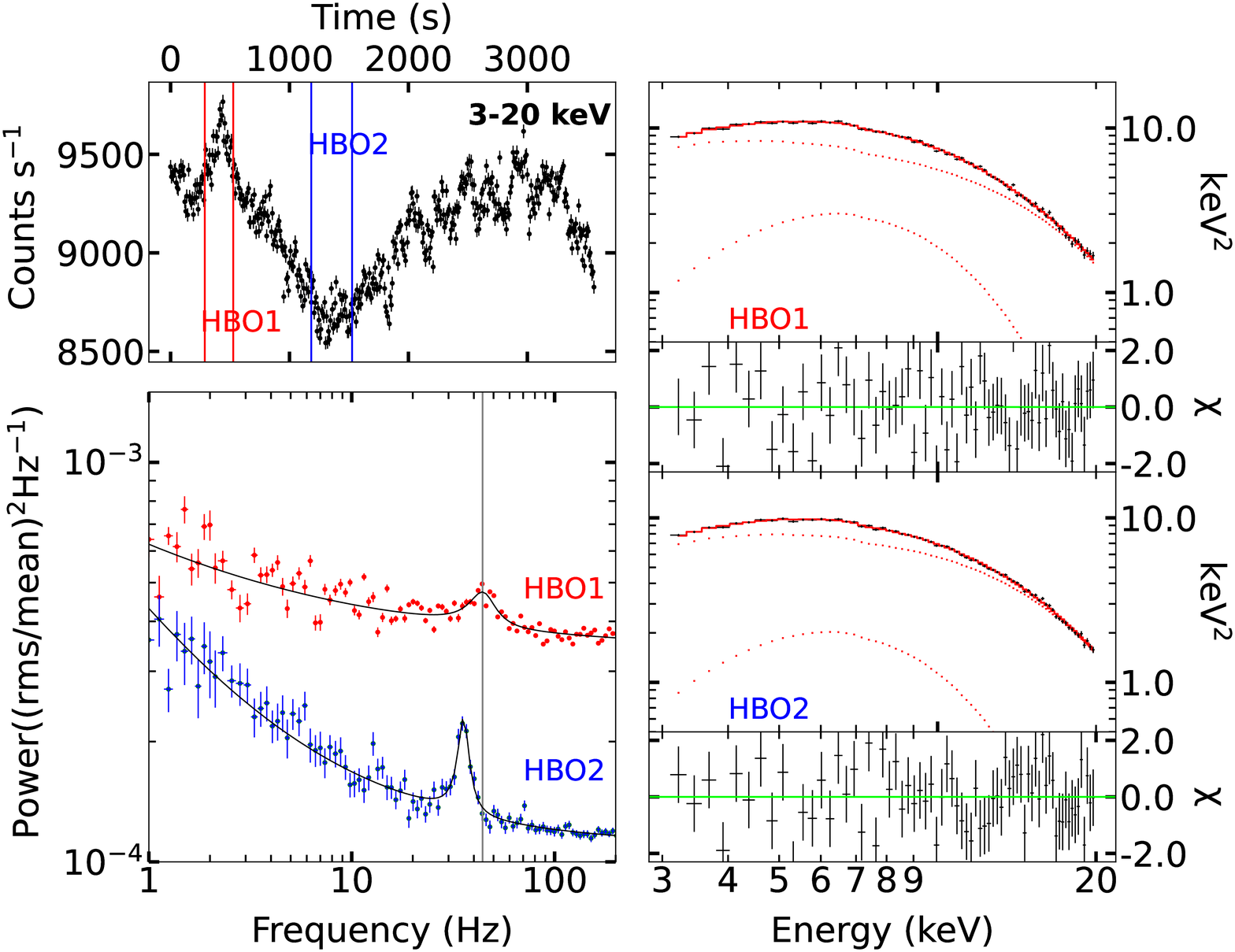}
\caption{Segment A}
\label{fig:subim1}
\end{subfigure}

\begin{subfigure}[b]{0.5\textwidth}
\includegraphics[width=9.3cm, height=7.5cm, angle=0]{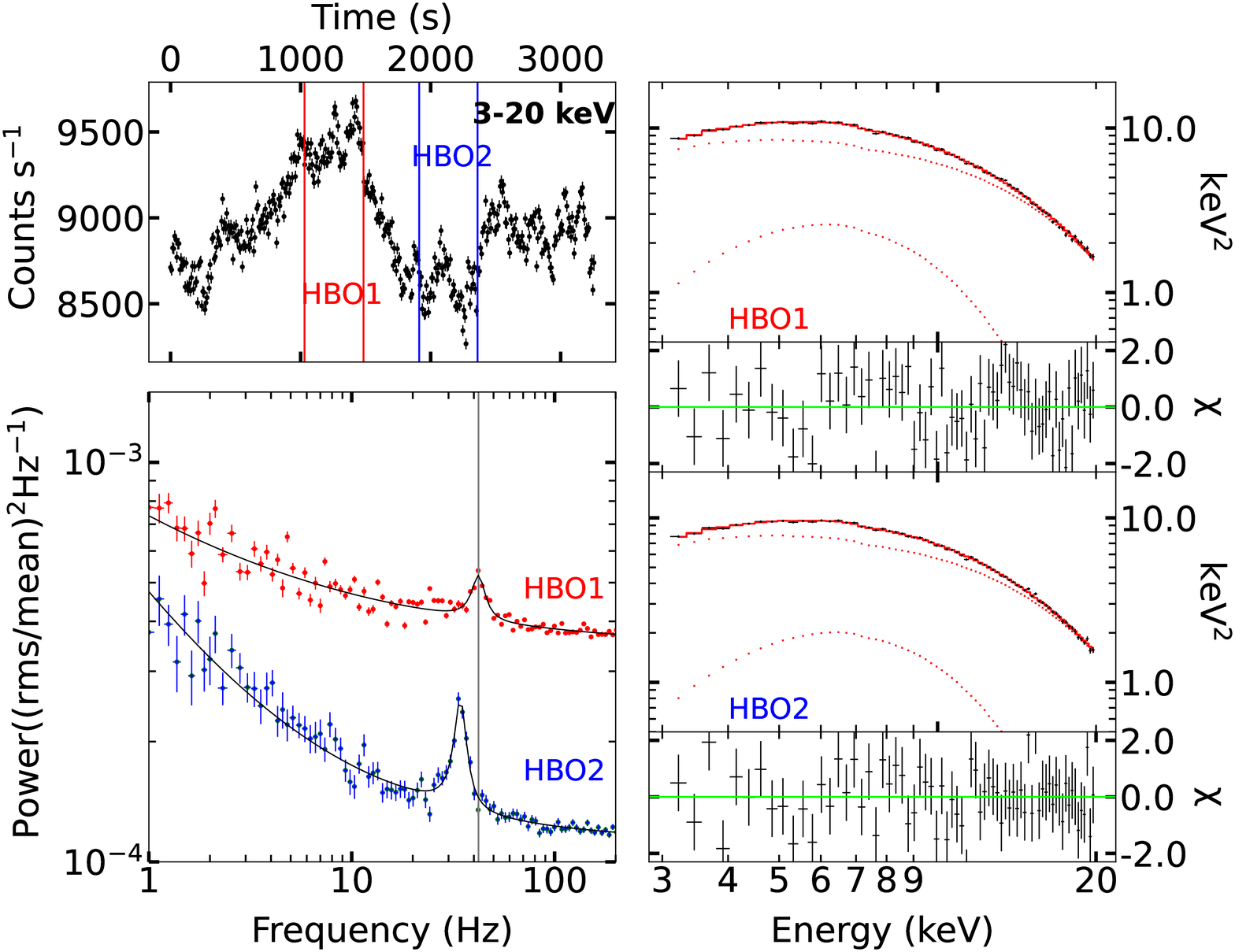}
\caption{Segment B}
\label{fig:subim2}
\end{subfigure}
\end{tabular}

\begin{tabular}{ c @{\quad} c }
\begin{subfigure}[b]{0.5\textwidth}
\includegraphics[width=9.3cm, height=7.5cm, angle=0]{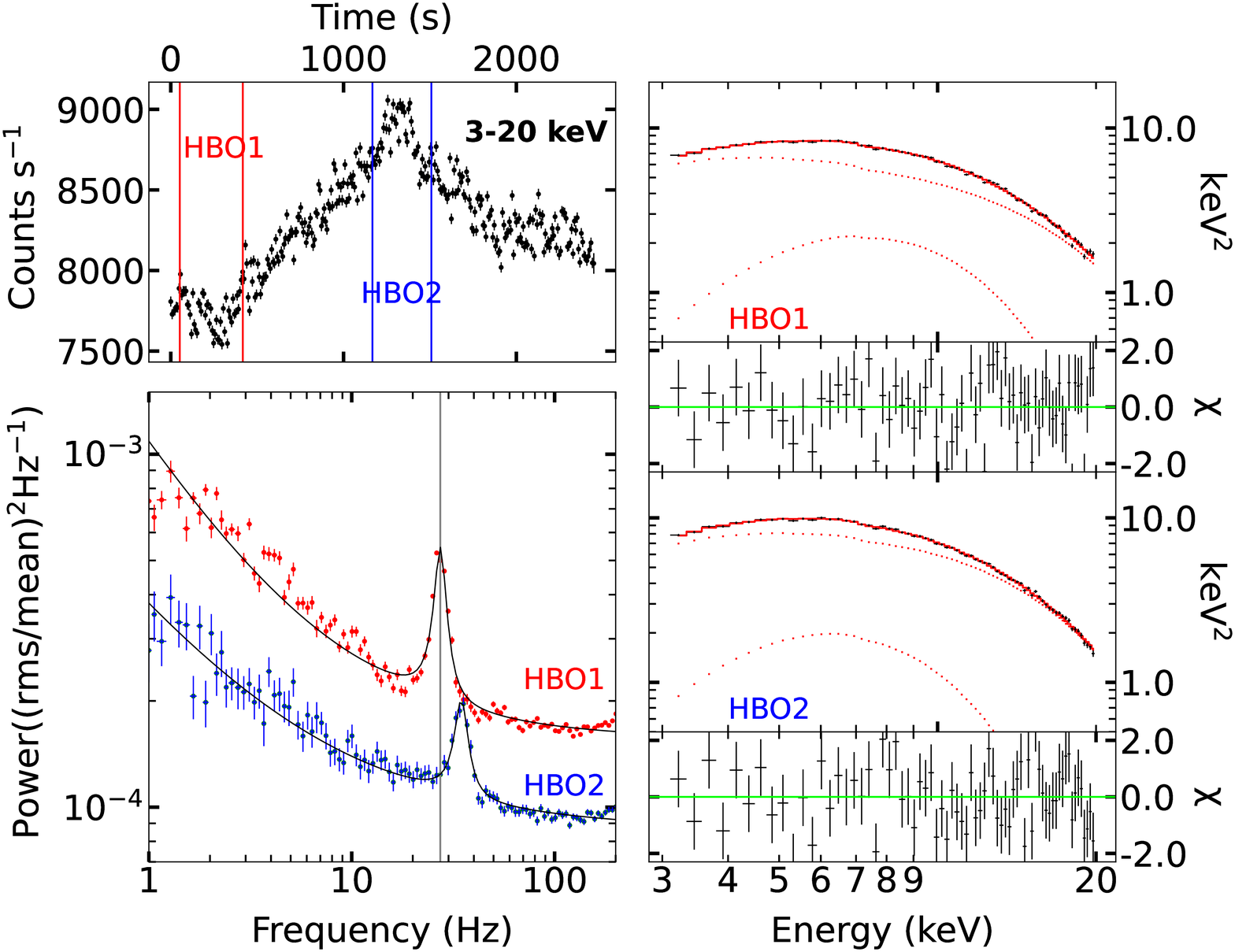}
\caption{Segment C}
\label{fig:subim3}
\end{subfigure}

\begin{subfigure}[b]{0.5\textwidth}
\includegraphics[width=9.3cm, height=7.5cm, angle=0]{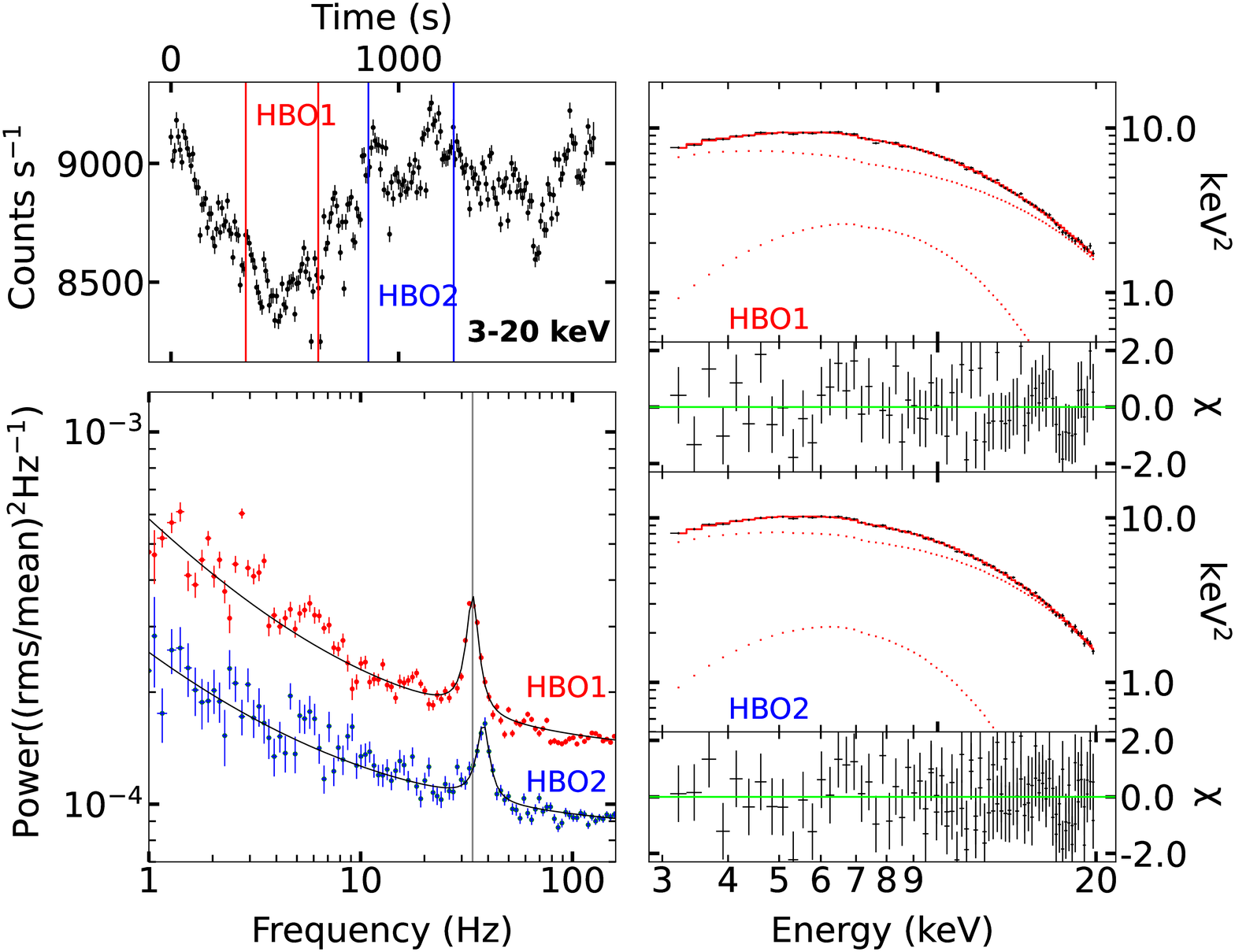}
\caption{Segment D}
\label{fig:subim4}
\end{subfigure}
\end{tabular}
\end{figure*}

\begin{figure*}
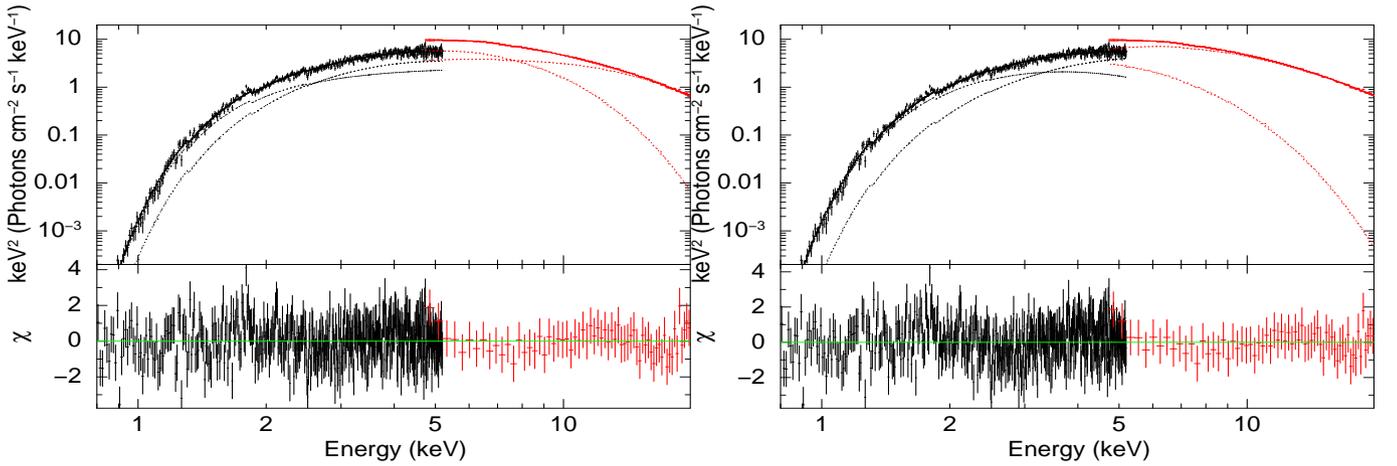

\centering
\caption {In each figure, the unfolded spectrum of GX 5-1 for p1: 0.8-5.2 keV (SXT) in black, 4.8-20 keV (LAXPC) in red colour along with their model components are shown. The lower panel shows the residuals of the spectral fit.}
\begin{tabular}{ c @{\quad} c }
\begin{subfigure}[b]{0.5\textwidth}
     \centering
     \includegraphics[width=6cm, height=9cm, angle=270]{bbody+comptt.eps}
     \caption{ The spectrum shown is unfolded using the model bbody+CompTT}
     \label{fig:subim1}
\end{subfigure}

\begin{subfigure}[b]{0.5\textwidth}
   \centering
  \includegraphics[width=6cm, height=9cm, angle=270]{diskbb+comptt.eps}
  \caption{The spectrum shown is unfolded using the model diskbb+CompTT}
  \label{fig:subim3}
\end{subfigure}

\end{tabular}

\begin{tabular}{ c @{\quad} c }
\begin{subfigure}[b]{0.5\textwidth}
    \centering
    \includegraphics[width=6cm, height=9cm, angle=270]{bbody+diskbb+powerlaw.eps}
    \caption{ The spectrum shown is unfolded using the model bbody+diskbb+powerlaw}
    \label{fig:subim4}
\end{subfigure}

\end{tabular}
\end{figure*}

\section {SPECTRAL FITS FOR LAG OBSERVATIONS ASSOCIATED TO HBO VARIATIONS}

In four observations where lags were detected, the spectral analysis was performed to constrain the responsible for causing these lags. We noted that during these observations HBO centroid frequencies varied as discussed above. For some sections, we fitted only the LAXPC 3-20 keV spectra as simultaneous SXT spectra were lacking. For the entire spectral analysis, LAXPC 10 data was used. Initially we used the model tbabs(diskbb+cutoffpl) where N$_{H}$ was fixed at 2.75 $\times$ 10$^{22}$ cm${^2}$. It was noticed that disc normalization has not varied during the HBO variations (Table 3). The cutoffpl $\Gamma$ has slightly varied eg. 2.40 to 1.99 for segment A as HBO varied from $\sim$44 Hz to $\sim$35 Hz, diskbb flux decreased (1.57 $\times$ 10$^{-8}$ erg cm$^{-2}$ s$^{-1}$ to 1.13 $\times$ 10$^{-8}$ erg cm$^{-2}$ s$^{-1}$) along with an increase in the cutoffpl flux (1.07 $\times$ 10$^{-8}$ erg cm$^{-2}$ s$^{-1}$ to 1.27 $\times$ 10$^{-8}$ erg cm$^{-2}$ s$^{-1}$). This indicates that during the presence of larger corona ($\nu$ $\sim$ 35 Hz), disc flux was low along with a higher cutoffpl flux. The cutoffpl $\Gamma$ varied from 2.82 to 1.99 for segment C, where HBO varied from $\sim$27 Hz to $\sim$35 Hz. The disc normalization has yet again not varied, disc flux decreased from 1.33 $\times$ 10$^{-8}$ erg cm$^{-2}$ s$^{-1}$ to 1.19 $\times$ 10$^{-8}$ erg cm$^{-2}$ s$^{-1}$ and cutoffpl flux increased from 0.78 $\times$ 10$^{-8}$ erg cm$^{-2}$ s$^{-1}$ to 1.24 $\times$ 10$^{-8}$ erg cm$^{-2}$ s$^{-1}$.  
Similar variations were observed in other segments too. We also tried another model i.e. bbody+CompTT. The hard component of the spectra was modelled with a Comptonization model i.e. CompTT model \citep{1994ApJ...434..570T}. This model has four parameters viz. electron temperature kT${_e}$,  soft photon temperature kT${_s}$,  optical depth $\tau$, and normalization. We noticed similar variations in fluxes whereas no noticeable variations were found in the model parameters (Table 4). In segments A and B, bbody and CompTT fluxes decreased as HBO frequency decreased, whereas bbody flux decreased along with an increase in CompTT flux in both C and D segments. This suggests that Compotnization region and/or BL were responsible for HBO frequency variations. Earlier Fourier frequency-resolved spectroscopic studies also indicated the BL is responsible for the 3-20 keV spectral continuum in these sources \citep{gilfanov2003boundary}.

We attempted a three-component model i.e. bbody+diskbb+power-law where the bbody model is assumed to be exhibited by a boundary layer on the NS. We observed a soft $\Gamma$ varying from 3.61 – 4.84 in segment A (Table 5). In segments, C and D as the HBO frequency increased, both bbody and disc fluxes increased along with a decrease in the power-law fluxes. In section B as the HBO frequency increased, we did not find significant variation in bbody and power-law fluxes along with an increase in the disc flux. In segment A, the bbody flux decreased, and diskbb and power-law fluxes increased as the HBO frequency increased. Moreover the total flux increased as the HBO frequency increased eg. for segment A it varied from 2.64 $\times$ 10$^{-8}$ erg cm$^{-2}$ s$^{-1}$ to 2.40 $\times$ 10$^{-8}$ erg cm$^{-2}$ s$^{-1}$ (see Table 5). It is found that disc flux varied along with the HBO frequency which is consistent with the truncated disc scenario i.e. as the HBO frequency increases, the disc flux also increases. But at the same time, we found a slight change in the inner disc radius of about 6 km in segments C and D, and no significant changes were noticed in the remaining segments (Table 5). Here it should be noted that we are interested in the relative changes in the R$_{eff}$. The power-law fluxes also decreased as the HBO frequency increased in segments C and D whereas such a scenario was not noticed in segment A.

We also attempted another way to ascertain the primary model parameter causing the lag where HBO frequency was found to vary. In this method, two spectra of a segment were fitted simultaneously using the above-mentioned model. One of the spectra was fixed at its best-fit parameters as obtained above and another one was tied to the first spectrum. This resulted in a high residual as expected eg. for segment A $\chi^2$/dof =10126.30 / 142  (Table 6). We later freed and untied the bbody parameter and noted the F-test value and the probability. Similarly, we united and freed the diskbb and power-law components independently. During this analysis, we noted that the diskbb model varied relatively more than other model parameters. The next model component whose variation is mandatory is the power-law model to explain the spectral variation among segments. For eg. in segment A, when the diskbb model is freed and united then it resulted in an F-test probability of about 1.71 $\times$ 10$^{-93}$ with an F-test value of 1410 whereas the power-law model requires an F-test probability of 2.63 $\times$ 10$^{-79}$ with an F-test value 763. This study indicates that disc and power-law models are the major components to cause the observed lags and HBO variations.

\section{SPECTRAL ANALYSIS FOR SIMULTANEOUS SXT+LAXPC DATA ACROSS THE HB AND NB} 
      
We also performed the spectral analysis of sections p1 and p6 using various models for SXT (0.8 to 5.2 keV) and LAXPC (4.8-20 keV) spectral data to constrain the variation in the physical component exhibiting the soft and hard components. \citet{bhulla2019astrosat} studied the same dataset using the model diskbb+Nthcomp and observed that the inner disk radius is almost stable across the HB and NB. We fitted the spectra with a model tbabs (bbody + CompTT) where N$_{H}$ was fixed at 2.75 $\times$ 10$^{22}$ cm$^{-2}$ (Figure 5, Table 7). We did not find any significant variations in any of the bbody and CompTT parameters across the p1 to p6. We found that the Comptonization region is characterized by low-temperature optically thick corona. We noted that bbody flux decreased from sections p1 and p6 i.e. from 2.03 $\times$ 10$^{-8}$ erg cm$^{-2}$ s$^{-1}$ to 1.02 $\times$ 10$^{-8}$ erg cm$^{-2}$ s$^{-1}$ and hard flux increased from 2.11 $\times$ 10$^{-8}$ erg cm$^{-2}$ s$^{-1}$ to 2.85 $\times$ 10$^{-8}$ erg cm$^{-2}$ s$^{-1}$ (Table 7). This indicates that boundary layer size is relatively larger in NB than in HB. 

In another model, we replaced the bbody model with the diskbb model assuming that the soft component is produced in the Keplerian portion of the accretion disc (Table 8). The inner disc temperature kT$_{in}$ was found to be varying from 1.16 to 1.27 keV and the inner disc radius to be around R$_{eff}$ $\sim$ 15--20 Km within the errorbars after applying the corrections. Similar to the above models, we did not find any variation in the Comptonization model parameters. Overall we suggest that the inner disc radius is close to the last stable orbit as noted from the arrived results.

We also employed a three-component model i.e. bbody+diskbb+power-law, assuming that bbody is arising from BL, diskbb is due to the soft photons arriving from the Keplerian portion of the disc, and the power-law component arising from the Comptonization region (Table 9). We noted that kT$_{bb}$ did not vary from p1 to p6 and was found to be around $\sim$ 1.20 keV, whereas kT$_{in}$ $\sim$ 2.47 keV to 2.89 keV but within error bars, such similar values were also noticed with PCA spectral fits \citep{sriram2012anti}. The inner disc radius (R$_{eff}$) was found to be low $\sim$ 5 km. The power-law indices was consistent to be around $\Gamma$ = 2.77$\pm$0.19 (p1) -- 2.68$\pm$0.35 (p6). The power-law component flux was relatively low in comparison to fluxes of diskbb and bbody components. The diskbb component fluxes from p1 to p6 were dominant, contributing $\sim$ 50—67 \% of total flux. It was noticed that bbody flux was decreasing from p1 to p6, suggesting that BL size is relatively smaller at the top of HB. The small size of BL at top of HB is consistent with the picture where the mass accretion rate is relatively low and hence material will not accumulate on the surface of NS, eventually forming a smaller BL. In GX 17+2, \citet{lin2012spectral} showed the BL size decreases its size at the lower vertex from a relatively larger size at the top of HB but how the BL size varies across the HB (top to upper vertex) in GX 5-1 or Cyg-type Z sources is not been studied extensively and hence more studies are needed in this direction. 

      \section{RESULTS AND DISCUSSION}
      
It is observed that in segments A and B, the HBO frequencies were decreasing along with total fluxes, where hard lag was observed in the former case and soft lag was observed in the latter (Table 3).  In segment A, the diskbb flux was found to be decreasing and cutoffpl flux was increasing, whereas in segment B, no change was observed in diskbb flux but cutoffpl flux was found to be increasing.  During the hard lag in segment A, the inner disc front was moving away from the NS based on the HBO shift, thereby decreasing the disc size along with an increase in the size of the Comptonizing region, which supports the corresponding changes in the respective component fluxes. In segment B, we noted a soft lag, which is not so significant in comparison with other segments (-50$\pm$40 s).

In segment C, HBO frequency was found to be increasing $\sim$ 27 Hz to $\sim$ 35 Hz with a hard lag of $\sim$ 702 s along with an increase in the total flux (Figure 4). The diskbb flux was decreasing from HBO1 to HBO2 sections along with an increase in the cutoffpl flux. This suggests that during the hard lag, the disc was moving away along with an increase in the size of the Comptonizing region as the HBO frequency was increasing. But in a truncated accretion disc scenario, as the QPO frequency increases the Comptonizing region size decreases, which was not observed in this segment. If we rely on the picture of the HBO shift, then the disc must have moved in. These observed changes can be explained if the corona is extended in the vertical direction, probably a jet-like structure which can explain both increase in the cutoffpl component flux as well as a decrease in the disc flux because the base of the jet might cover some portion of the inner region of the disc. The possibility of the jet cannot be ruled out as the radio jet is observed to be present in the HB \citep{penninx1989non, 2000MNRAS.318..599B}. In segment D, the HBO was found to vary from $\sim$ 34 Hz to $\sim$ 38 Hz associated with a soft lag of $\sim$ -- 192 s with a similar variation in the total flux. Moreover, the diskbb flux was found to decrease along with an increase in the cutoffpl flux. Again a vertically extended corona or a jet can explain the observed variations in the flux, as discussed in the case of the segment D scenario.

In case of the bbody+CompTT model, CompTT flux has not changed whereas bbody flux has decreased as HBO frequency decreased in the case of segment A (Table 4). In segments C and D, bbody fluxes decreased whereas CompTT flux decreased in the former and increased in later segments as HBO frequency increased from HBO1  to HBO2 sections.  In segment C, during the hard lag, the BL size is decreased along with an increase in the size of the Comptonizing region whereas similar variations were noted in segment D associated with a soft lag. It is clear from this study that BL is also responsible for observed lag as well as HBO frequency variations and variation in the size of BL is causing the observed lags.

In the three-component model (Table 5), we noticed that R$_{eff}$ and $\Gamma$ varied from HBO 1 to HBO 2 in segments C and D and no noticeable variation was observed in R$_{eff}$ in segment A. The bbody flux decreased along with an increase in diskbb and power-law fluxes as HBO frequency decreased from HBO1 to HBO2.  Whereas in segments C and D, bbody and diskbb fluxes were increased but power-law fluxes were noted to be decreasing as HBO frequencies were shifting to higher frequencies.  In both cases, the disc component increased due to an increase in the mass accretion rate, which is helping to increase the size of BL on the NS. At the same time, the increase in soft photons cools the Comptonizing region which is evident from the decrease in the power-law flux. So the lags detected in these sections suggest that there is a reduction in the size of the Comptonizing region. Perhaps all three physical structures are needed to vary in order to explain the detected lags.

On the other hand in segment A, as the HBO frequency decreased the disc has decreased in its size (based on the flux), which is well in agreement with the geometry of the truncated disc model. But the power-law flux assuming to arising from the Comptonizing region was found to be decreasing which should be increasing if the disc is moving out. In general, in the truncated disc model, the decrease in the QPO frequency is associated with the increase in the Comptonizing region flux (i.e. power-law flux). However, we noticed that BL size (based on the bbody flux) has increased as HBO frequency decreased and assuming that if BL is also responsible for some portion of the hard X-ray continuum, then the decrease in HBO frequency can be explained. It is possible that the BL is also somehow connected to HBO, however, the exact association is not known and more studies are required in this direction.

Overall it is evident that the observed lags are associated with the readjustment time scale of the inner region of the accretion disc and various components viz. Keplerian portion of the disc, BL and Comptonizing region are responsible for the lags detected. These results were also supported by the fact the HBO frequency changes during the detected lags.

\subsection{Size of Inner hot flow using Lense-Thirring Precession}
It is difficult to constrain the size of the corona in neutron star Z sources. However, HBO provides a means by which one can know the size of corona, assuming to be producing HBO. Here we use the Lense-Thrring precession model developed by \citet{ingram2009low}, where inner hot rigid accretion flow is misaligned with the compact object spin. It has been strongly argued that the LF QPOs are produced in the Comptonized region rather than associated with the disc \citep{sobolewska2006spectral}. In this model, the inner hot flow is vertically precessing like a rigid body whereas its outer region is truncated by the Keplerian disc and the inner radius is approximately constant close to the NS  \citep{ingram2010physical}. The mass distribution in the hot flow, which may be parameterized by $\zeta$, the radial dependence of the surface density, $\sum$=$\sum_{o}$(r/r$_i$)$^{-\zeta}$ \citep{ingram2009low, ingram2010physical, fragile2007global}. The frequency of the LF QPO is therefore estimated to be
  \begin{equation}
   \nu_{prec} = \frac{(5-2\zeta)}{\pi (1+2\zeta)}\frac{a_{*}[1-(r_i/r_o)^{1/2+\zeta}]}{{r_o}^{5/2-\zeta}{r_i}^{1/2+\zeta}[1-(r_i/r_o)^{5/2-\zeta}]}\frac{c}{R_g}
  \end{equation}
  
  While simulation indicates $\zeta$ $\sim$ 0 \citep{fragile2007global} and a similar value can be adopted for neutron stars, which may result in a situation where the flow becomes increasingly concentrated on the neutron star surface due to increase in the accretion rate. However, assuming $r_i$ and $r_o$ are variable with frequency, the observed HBO variations can be calculated.
  
Here we calculated the outer radius of the inner hot flow using the above equation, the inner disc radius of the inner hot flow $r_i$ $\sim$ 4.5 $R_g$  (9.29 km) was fixed \citep{ingram2010physical}. We got the values $r_o$ $\sim$ 15.45 km (7.5 $R_g$ ) (from the Table 5, $R_{eff}$ = 11.79 km)   for 44.14 Hz HBO. $r_o$ $\sim$ 17.53 km ( 8.49 $R_g$ ) (from the Table 5, $R_{eff}$ = 10.75 km) for 35.56 Hz HBO. $r_o$ $\sim$ 20.23 km (9.8 $R_g$) ( from the Table 5, $R_{eff}$ = 7.43 km ) for 27.33 Hz HBO.  These $r_o$ values are different from the observed apparent inner disc radius $\sim$ 8 km for 27 Hz (Table 5). However similar values of R$_{in}$ were noticed with the model diskbb+CompTT (Table 8).  It is known that it is difficult to estimate the true inner disc radius as the apparent inner disc radius must undergo a few corrections (see section 6.3).

  \subsection{Coronal height estimation}
  
We can estimate the coronal height using the equation based on the measured lags \citet{sriram2019constraining}. This equation was derived assuming that the observed lags are due to the Corona/sub-Keplerian flow readjustment timescales in the inner region of the accretion disc. Since in this source, the inner disc radius was found to be always close to the last stable orbit that is well in agreement with other studies \citep{homan2018absence}, hence the only varying structure would be the Comptonizing region in the inner region. As a result, we may constrain the coronal size using the equation below.
  
  \begin{equation}
  \ H_{corona}=\Bigg(\frac{t_{lag}\dot{m}}{2\pi R_{disc} H_{disc} \rho}-R_{disc}\Bigg)\times \beta  \quad cm
 \end{equation}

            Here is   
             \[H_{disc} = 10^{8}\alpha^{-1/10} \dot{m}_{16}^{3/20} R_{10}^{9/8} f^{3/20} \quad cm\]
        \[\rho = 7\times 10^{-8} \alpha^{-7/10} \dot{m}_{16}^{11/20} R_{10}^{-15/8} f^{11/20} \quad g \ cm^{-3} \]
        \[f = (1-(R_s/R)^{1/2})^{1/4}\]
        and \[\beta = v_{corona}/v_{disc}\] \citep{shakura1973black, sriram2019constraining}.
        
For the light curve segments with CCF lags, the coronal height was calculated. Based on the radius determined from the spectral model (see Table 8), the disc radius R$_{disc}$ was taken to be 20 km, and $\beta$ was taken to be 0.05–0.1. \citep{manmoto1997spectrum, pen2003fate, mckinney2012general}.
For each segment, an average R$_{disc}$ value of 20 km was chosen, and $\dot{m}$ was calculated using the luminosity, L = GM$\dot{m}$/R assuming M =1.4 M$_{\odot}$ and $\alpha$ was taken to be 0.1.

The coronal height estimated to be 27-35 km ($\beta$ = 0.05) and 55-71 km ($\beta$ = 0.1) for a 341 s lag. For 335 s lag, the height was estimated to be 25-36 km ($\beta$ = 0.05) and 50-73 km for $\beta$ = 0.1. The coronal height is constrained to 9-23 km for $\beta$= 0.05 and 18-46 km for $\beta$ =0.1  height for a lag of 702 seconds. For 183 s lag, the coronal height estimated to be 58-73 km ($\beta$ = 0.05) and 116-147 km ($\beta$ = 0.1).        
For the soft lag -198 s coronal height was estimated around 16-19 km ($\beta$=0.05) and 32-38  ($\beta$=0.1). Similarly for -230 s and -157 s lags, coronal heights were determined to be 18-23 km ($\beta$=0.05) , 36-46 km ($\beta$=0.1) and 9-18 km ($\beta$=0.05) , 19-36 km ($\beta$=0.1) respectively for soft lags.      

As these lags might be caused due to the change in the size of the boundary layer (BL), we calculated its size using the equation \citet{ popham2001accretion},

 \begin{equation}
  \ log(R_{BL}-R_{Ns}) \sim 5.02+0.245\Bigg[log \bigg(\frac{\dot{M}}{10^{-9.85} M_{\odot} yr^{-1}}\bigg)\Bigg]^{2.19}
 \end{equation} 
here $\dot{M}$ obtained from the equation $L=\frac{GM\dot{M}}{R}$ with M = 1.4$M_\odot$ and R = 10 km. Luminosity is obtained from spectral fit (Table 8). The values of R$_{BL}$ were determined to be 36 km, 55 km, 52 km, 48 km, 42 km and 32 km for sections p1 to p6 respectively.

Since the Z source radiate close to or slightly above the Eddington luminosity, the inner region of the disc modifies its shape. Hence we also calculated the spherization radius of the puffed-up disc due to radiation pressure in the inner region of the accretion disc \citep{ding2011nature}.

 \begin{equation}
  \ R_{sp} \sim \frac{9}{4} \times 10^{6} m \dot{M}_{disc,Edd} \quad cm
 \end{equation}
 
 Here $\dot{M}_{disc, Edd}$ is the disc accretion rate in units of Eddington accretion rate, m is the mass of a compact object in units of solar mass. For a neutron star mass of 1.4 $M_{\odot}$, the above equation becomes 
 
  \begin{equation}
  \ R_{sp} \sim 32 \dot{M}_{disc,Edd} \quad km
 \end{equation}
 
Using the Eddington luminosity 3.8 $\times$ 10$^{38}$ erg s$^{-1}$ \citep{2003A&A...399..663K, ding2011nature}, the Eddington mass accretion rate is found to be 2.045 $\times$ 10$^{18}$ g s$^{-1}$. From the above equation, R$_{sp}$ was estimated to be about $\sim$ 34 km assuming the total flux observed in section p1 (Table 8). For section p2, R$_{sp}$ was found to be $\sim$ 41 km and 31 km for section p6. 

\subsection{Constraining the Inner disc radius}
Based on the spectral fits obtained from the diskbb+compTT model, we estimated the inner disc radius using N = (R$_{in}$/D$_{10}$)$^2$cos(i) where N is diskbb model normalization \citep{mitsuda1984energy}, where D$_{10}$ is the distance to the source in units of 10 kpc and R$_{in}$ is the inner disc radius. We obtained R$_{in}$ to be 19.35 km, 18.99 km, 17.52 km, 15.84 km, 17.24 km and 15.95 km using an inclination angle i of 60$^{\circ}$ and a distance 9 kpc for different sections (Table 8) \citep{christian1997survey,homan2018absence}. For these apparent inner disc radii, we applied the correction factors in terms of spectral hardening ($\kappa$) and inner boundary condition ($\zeta$) to estimate the true inner disc radii. \citet{shimura1995spectral} found a $\kappa$ value of 1.7–2.0 and a $\zeta$ value of 0.41 \citep{kubota1998evidence}. As a result, the true radii values have slightly increased R$_{eff}$ can  be written as $\kappa$$^2$$\zeta$R$_{in}$ \citep{kubota2001observational}. The values of R$_{eff}$ for different sections were found to be 22.93 km, 22.50 km, 20.76 km, 18.77 km, 20.43 km and 18.89 km (Table 8). As a result, the disc seems to be near the NS and stationary along the Z-track as the observed variations are within the error bars.

 We also calculated the inner disc radius from the Relativistic Precession Model (RPM) based on the observed HBOs that were found to be varying from $\sim$ 44 Hz to $\sim$ 27 Hz. The HBOs and kHz QPOs are related to the Keplerian orbital motion, nodal precession and precession of periastron of the perturbed orbit in the RPM \citep{stella1999khz, stella1999correlations}. The inner disc radius was estimated using the following equation \citep{di1999quasi,sriram2019constraining}
    
    \begin{equation}
       \frac{R_{in}}{R_g}\leq 27 \nu^{-0.35}\bigg(\frac{M}{2 M_{\odot}}\bigg)^{-2/3}
    \end{equation}
    we found R$_{in}$ to be $\sim$9 R$_g$ ($\sim$18 km) for $\sim$44 Hz HBO, and R$_{in}$ to be $\sim$ 11 R$_g$ ($\sim$22km) for $\sim$27 Hz HBO. These values are in good agreement with the the values of true inner disc radii obtained from the spectral fits (Table 8).

We also invoked the Transition Layer Model (TLM) \citep{osherovich1999kilohertz, osherovich1999interpretation, titarchuk1999correlations} to estimate the inner disc radius using the relation between HBO and kHz QPO in Z sources. The equation from \citet{sriram2019constraining}
 \begin{equation}
   \frac{R_{in}}{R_g}\leq 220\Bigg[\bigg(\frac{\Omega}{\pi}\bigg)^{-2/3} \Bigg(\bigg(\frac{\Omega sin \delta}{\pi \nu_{HBO}}\bigg)^{2}-1\Bigg)^{1/3}\Bigg]\bigg[\frac{M}{10 M_{\odot}}\bigg]^{-2/3}
 \end{equation}  
 here we took $\delta$ = 5.80 and $\Omega/2\pi$ = 340 Hz \citep{wu2001testing} is the angle between rotational angular velocity $\Omega$ and normal to the Keplerian oscillation for GX 5-1 \citep{wu2001testing}.
We found R$_{in}$ to be $\sim$ 12 R$_g$ ($\sim$ 25 km) for HBO frequency $\sim$ 44 Hz, and R$_{in}$ to be $\sim$ 19 R$_g$ ($\sim$ 38 km) for $\sim$ 27 Hz.

  \section{CONCLUSION}   
Below are the conclusions of the present work.

1. For the first time, we report the cross-correlation functions between soft (SXT, 0.8-2.0 keV) and hard (LAXPC 10-20 keV, 16-40 keV) of the source GX 5-1. Anti-correlated soft lags were detected in  the order of a few tens to hundreds of seconds. We also report the hard and soft anti-correlated lags of the order a few tens to hundreds of seconds using 3-5 keV and 16-40 keV energy bands using LAXPC data only.

2. During the detected lag , the HBO centroid frequency was found to vary from higher to lower frequency and vice-versa in different observations. These variations on HB strongly indicate that the inner disc structure has changed during the detected lags. The spectral analysis unveiled that the inner disc radius varied on two occasions (Table 5) whereas other model parameters remain constant. Assuming that the HBOs are producing in the Comptonizing region, we suggest that the corona size is varying during the detected lags.

3. We interpreted that lags as the readjustment time scale during which the inner disc region is adjusted. Assuming the disc is at the last stable orbit as indicated by the spectral fits, we suggest that during the observed lags the corona size has changed. However, we do not rule out the possibility that both BL and corona are readjusted during the detected lags. We estimated the size of the corona to be 15-40 km for observed lags from the various models. We also found that corona size is of the order of 8-25 km based on the detected HBOs and BL size to be 32-55 km. All these results suggest that a few tens of km of the corona is needed to explain lags, HBOs and spectral variation on the HB branch.

4. We fitted the SXT and LAXPC data covering a spectral domain of 0.8-20 keV using different model components. We found that the inner disc radius is close to the last stable orbit. Such a similar configuration is also been observed using NuStar data also \citep{homan2018absence}. Based on the results, we found that BL size is relatively smaller at the top of the HB, however, more observations are needed especially covering the low energy spectral domain (< 3 keV).

Future high-resolution X-ray spectral and timing observations would give more details about the structural and geometrical variations during the lags in Z sources and the role of the BL along with the Comptonization region in modulating the HBO can be understood more vividly.

\begin{table*}
\centering
\caption{Results of the fit to the cross-correlation functions of LAXPC and SXT light curves of observation ID G06-114T01-9000001056,  2017-02-26 17 34 55 (Start time (UT), 2017-02-27 06 45 17 (Stop time (UT)). The first column represents the section name (see Figure 3), the second and third columns represent the start and stop time and exposure time are shown in the fourth column. The fifth and sixth columns show the correlation coefficient (CC) and detected lags. Here sections A and D the lags were detected in the energy bands between 0.8--2 keV and 10--20 keV and in sections B, C and E energy bands were 0.8-2 keV and 16-40 keV. The remaining sections F to P energy bands were 3-5 keV and 16-40 keV.}
\begin{tabular} {c c c c c c}
\hline
\hline
Section & Start time(UT) &  Stop time(UT) & Exposure time (s) & CC & Lags  (s) \\
\hline
\hline
A &   2017-02-26 18 25 31  &   2017-02-26 18 43 13 & 1062 &  -0.35$\pm$0.04 & -198$\pm$17   \\

B & 2017-02-26 18 25 31  &    2017-02-26 18 43 13  & 1062 & -0.34$\pm$0.05 & -230$\pm$27  \\ 

C &  2017-02-26 20 02 58 &   2017-02-26 20 20 58   & 1080 & -0.39$\pm$0.20 & -56$\pm$39   \\ 

D  &  2017-02-26 21 40 25  &    2017-02-26 21 57 48  & 1043 &  -0.40$\pm$0.04 &  -256$\pm$24  \\

E  & 2017-02-26 21 40 25 &  2017-02-26 21 57 48  & 1043 & -0.39$\pm$0.05 &  -235$\pm$28   \\

F & 2017-02-26 18 25 53 &  2017-02-26 19 15 03  & 2950 & 0.76$\pm$0.02 &  1.67$\pm$4   \\ 

G & 2017-02-26 20 03 17 &  2017-02-26 20 33 37  & 1820 & -0.43$\pm$0.15 &  341$\pm$42   \\

H & 2017-02-26 21 40 45 &  2017-02-26 22 00 35  & 1190 & -0.65$\pm$0.05 &  -63$\pm$17   \\

I & 2017-02-26 22 24 21 &  2017-02-26 22 40 01 & 940 & -0.45$\pm$0.14 &  19$\pm$26   \\

J & 2017-02-26 23 18 07 &  2017-02-26 23 54 57  & 2210 & -0.35$\pm$0.05 &  335$\pm$60   \\

K & 2017-02-26 23 54 47 &  2017-02-27 0 17 57 & 1390 &  0.43$\pm$0.08 &  -157$\pm$45   \\

L & 2017-02-27 1 08 13 &  2017-02-27 1 44 43  & 2190 &  0.47$\pm$0.05 &  -59$\pm$40   \\

M & 2017-02-27 2 45 03 &  2017-02-27 3 23 18 & 2295 & -0.33$\pm$0.06 &   183$\pm$73   \\

N & 2017-02-27 3 21 43 &  2017-02-27 3 32 33 & 650 &  0.53$\pm$0.14 &  -92$\pm$33   \\

O  & 2017-02-27 4 29 05 &  2017-02-27 5 09 25 & 2420 & -0.48$\pm$0.04 &  702$\pm$80  \\

P & 2017-02-27 6 13 39  &  2017-02-27 6 44 09  & 1830 & -0.46$\pm$0.04 &  -194$\pm$27   \\
\hline
\end{tabular}
\end{table*}

\begin{table}
\centering
\caption{Detected HBOs along with Lorentzian parameters for the respective segments. The second column is the centroid frequency of HBO, the third column is the FWHM of HBO and the fourth column is the quality factor, Q= $\nu$ / $\Delta\nu$.}
\begin{tabular}{cccccccccccccc}
\hline  \hline
   Segment  & $\nu$ (Hz) & $\Delta\nu$ (Hz) & Q-factor & $\chi^{2}$/dof \\
\hline  \hline
 A \\\cline{1-1}

HBO1      &44.14$^{+1.97}_{-2.00}$ &15.91$^{+5.22}_{-7.39}$ & 2.77  & 89/99\\
 
HBO2      &35.56$^{+0.33}_{-0.33}$ &5.02$^{+0.77}_{-0.88}$ & 7.08 & 112/99\\

 B \\\cline{1-1}

HBO1      &42.15$^{+0.75}_{-0.70}$ &7.56$^{+2.13}_{-2.91}$ & 5.57 & 105/99\\
 
HBO2      &34.31$^{+0.25}_{-0.26}$ &5.31$^{+0.67}_{-0.76}$ & 6.46  & 119/99\\

 C \\\cline{1-1}

HBO1      &27.33$^{+0.19}_{-0.20}$ &4.07$^{+0.33}_{-0.36}$ & 6.71  & 154/108\\
 
HBO2      &34.71$^{+0.32}_{-0.32}$ &5.91$^{+0.73}_{-0.81}$ & 5.87  & 145/108\\

 D \\\cline{1-1}

HBO1      &33.76$^{+0.25}_{-0.27}$ &4.81$^{+0.58}_{-0.65}$ & 7.01  & 141/116\\ 

HBO2      &37.93$^{+0.38}_{-0.37}$ &5.67$^{+0.97}_{-1.15}$ & 6.68  & 152/116\\

\hline
\end{tabular}
\end{table}


\begin{table*}
\centering
\caption{Best-fit parameters of the spectra of different segments using the model  $\tt  tbabs(diskbb+cutoffpl)$. $k$T$_{in}$ is the inner disc temperature and N$_{dbb}$ is the normalization of the diskbb model. $\Gamma$ is the power-law photon index of the cutoffpl model. E$_{cpl}$ is the e-folding energy of exponential roll-off. N$_{cpl}$ is the normalization of cutoffpl. Flux is in the units of 10$^{-08}$ ergs cm$^{-2}$ s$^{-1}$. All the errorbars quoted are 90\% confidence level.}
\begin{tabular}{ccccccccccccccc}
\hline \hline
\multicolumn{1}{c}{Parameter} & \multicolumn{2}{c}{\text{Segment A}} && \multicolumn{2}{c}{\text{Segment B}} && \multicolumn{2}{c}{\text{Segment C}} && \multicolumn{2}{c}{\text{Segment D}} \\ [0.5ex]
\cline{2-3}  \cline{5-6} \cline{8-9} \cline{11-12}
\textbf{} & {HBO1} & {HBO2} & & {HBO1} & {HBO2} && {HBO1} & {HBO2} && {HBO1} & {HBO2} \\
\hline \hline

$k$T$_{in}$ (keV) & 2.47$^{+0.04}_{-0.05}$ & 2.58$^{+0.15}_{-0.06}$ && 2.47$^{+0.07}_{-0.05}$ & 2.62$^{+0.10}_{-0.05}$ && 2.72$^{+0.06}_{-0.03}$ & 2.64$^{+0.17}_{-0.07}$  && 2.59$^{+0.05}_{-0.04}$  & 2.57$^{+0.13}_{-0.06}$ \\[1.0ex]

N$_{dbb}$ & 28.30$^{+4.79}_{-7.12}$ & 17.16$^{+5.41}_{-7.96}$ && 23.73$^{+5.75}_{-9.08}$ & 17.92$^{+4.20}_{-6.04}$ && 15.86$^{+1.28}_{-3.07}$ & 16.52$^{+5.22}_{-7.41}$ && 20.44$^{+3.45}_{-4.13}$  & 18.76$^{+5.53}_{-8.32}$  \\[1.0ex]

$\Gamma$ & 2.40$^{+0.58}_{-0.53}$ & 1.99$^{+0.50}_{-0.47}$ && 1.97$^{+0.50}_{-0.47}$ & 2.20$^{+0.49}_{-0.47}$ && 2.82$^{+0.10}_{-0.49}$ & 1.99$^{+0.50}_{-0.47}$ && 2.57$^{+0.40}_{-0.44}$  & 1.99$^{+0.51}_{-0.48}$ \\[1.0ex]

E$_{cpl}$ & 13.58$^{+9.28}_{-6.45}$ & 8.56$^{+3.79}_{-2.83}$ && 8.51$^{+3.30}_{-2.76}$ & 10.36$^{+5.87}_{-3.81}$ && 41.26$^{+34.20}_{-31.52}$ & 8.33$^{+3.9}_{-2.73}$ && 22.23$^{+14.79}_{-6.58}$  & 8.55$^{+3.82}_{-2.84}$ \\[1.0ex]

N$_{cpl}$ & 14.13$^{+6.59}_{-4.17}$ & 10.32$^{+3.40}_{-1.94}$ && 10.53$^{+3.42}_{-2.22}$ & 11.42$^{+3.87}_{-2.40}$  && 15.29$^{+2.77}_{-4.28}$ & 10.33$^{+3.39}_{-1.96}$ && 13.76$^{+5.03}_{-3.15}$  & 10.53$^{3.51}_{-2.00}$  \\[1.0ex]

Flux$_{dbb}$ & 1.57 & 1.13 && 1.28 & 1.26 && 1.33 & 1.19 && 1.38 & 1.20  \\ [1.0ex]

Flux$_{cpl}$ & 1.07 & 1.27 && 1.35 & 1.11 && 0.78 & 1.24 && 0.96 & 1.29 \\[1.0ex]

Flux$_{total}$ & 2.64 & 2.40 && 2.63 & 2.37 && 2.11 & 2.43 && 2.34 & 2.49   \\  [1.0ex]

$\chi^{2}$/dof & 90/66 & 80/66 && 88/66 & 75/66 && 88/66 & 86/66 && 76/66 & 71/66 \\[1.0ex]

$\nu$ (Hz) & 44.14$^{+1.97}_{-2.00}$ & 35.56$^{+0.33}_{-0.33}$ && 42.15$^{+0.75}_{-0.70}$ & 34.31$^{+0.25}_{-0.26}$ && 27.33$^{+0.19}_{-0.20}$ & 34.71$^{+0.32}_{-0.32}$ && 33.76$^{+0.25}_{-0.27}$ &   37.93$^{+0.38}_{-0.37}$ \\[1.0ex] 

lag (s) & \multicolumn{2}{c}{\text{335$\pm$60}} & & \multicolumn{2}{c}{\text{-59$\pm$40}} & & \multicolumn{2}{c}{\text{702$\pm$80}} & & \multicolumn{2}{c}{\text{-194$\pm$27}} \\[1.0ex]

\hline
\end{tabular}
\end{table*}

\begin{table*}
\centering
\caption{Best-fit parameters of the spectra of different segments using the model $\tt tbabs(bbody+CompTT)$. $k$T$_{bb}$ is the black body temperature and N$_{bb}$ is the normalization of the blackbody component. T$_{0}$ is the Input soft photon temperature of the CompTT model. $k$T$_e$ is the plasma temperature. $\tau$ is the plasma optical depth. N$_{cptt}$ is the normalization of CompTT. Flux is in units of 10$^{-08}$ ergs cm$^{-2}$ s$^{-1}$.}
\begin{tabular}{ccccccccccccccc}
\hline \hline
\multicolumn{1}{c}{Parameter} & \multicolumn{2}{c}{\text{Segment A}} && \multicolumn{2}{c}{\text{Segment B}} && \multicolumn{2}{c}{\text{Segment C}} && \multicolumn{2}{c}{\text{Segment D}} \\ [0.5ex]
\cline{2-3}  \cline{5-6} \cline{8-9} \cline{11-12}
\textbf{} & {HBO1} & {HBO2} & & {HBO1} & {HBO2} && {HBO1} & {HBO2} && {HBO1} & {HBO2} \\
\hline \hline

$k$T$_{bb}$ (keV) & 1.59$^{+0.11}_{-0.10}$ & 1.52$^{+0.12}_{-0.11}$ && 1.49$^{+0.10}_{-0.11}$ & 1.58$^{+0.13}_{-0.11}$  && 1.79$^{+0.10}_{-0.08}$  & 1.53$^{+0.13}_{-0.12}$  && 1.68$^{+0.11}_{-0.16}$ & 1.51$^{+0.12}_{-0.11}$ \\[1.0ex]

N$_{bb}$ & 0.082$^{+0.019}_{-0.022}$ & 0.055$^{+0.011}_{-0.013}$ && 0.071$^{+0.011}_{-0.013}$ & 0.055$^{+0.012}_{-0.015}$ && 0.059$^{+0.017}_{-0.016}$  & 0.054$^{+0.012}_{-0.014}$  && 0.070$^{+0.018}_{-0.020}$  & 0.059$^{+0.011}_{-0.014}$ \\[1.0ex]

$k$T$_{0}$ (keV) & 0.40$^{+0.08}_{-0.10}$ & 0.35$^{+0.09}_{-0.10}$ && 0.36$^{+0.10}_{-0.11}$ & 0.37$^{+0.09}_{-0.11}$ && 0.35$^{+0.08}_{-0.12}$ & 0.37$^{+0.08}_{-0.11}$ && 0.42$^{+0.11}_{-0.13}$  & 0.37$^{+0.09}_{-0.11}$ \\[1.0ex]

$k$T$_e$ (keV) & 3.16$^{+0.24}_{-0.46}$ & 3.03$^{+0.21}_{-0.14}$ && 3.00$^{+0.19}_{-0.13}$ & 3.05$^{+0.15}_{-0.23}$ && 3.51$^{+0.32}_{-0.53}$ & 2.99$^{+0.14}_{-0.20}$ && 3.46$^{+0.31}_{-0.69}$  & 3.01$^{+0.14}_{-0.20}$ \\[1.0ex]

$\tau$ & 4.49$^{+0.71}_{-0.52}$ & 4.91$^{+0.36}_{-0.29}$ && 4.91$^{+0.46}_{-0.35}$ & 4.86$^{+0.48}_{-0.38}$  && 4.16$^{+0.69}_{-0.54}$ & 5.01$^{+0.45}_{-0.36}$ && 4.21$^{+0.81}_{-0.54}$  & 4.96$^{+0.44}_{-0.36}$  \\[1.0ex]

N$_{cptt}$ & 3.87$^{+0.77}_{-1.13}$ & 4.09$^{+0.79}_{-1.26}$ && 4.37$^{+0.67}_{-1.30}$ & 3.81$^{+0.62}_{-1.10}$  && 3.26$^{+0.53}_{-0.98}$ & 3.90$^{+0.79}_{-1.23}$ && 2.91$^{+0.82}_{-0.98}$  & 3.98$^{+0.82}_{-1.20}$  \\[1.0ex]

Flux$_{bb}$ & 0.58 & 0.38 && 0.49 & 0.39 && 0.44 & 0.38 && 0.51 & 0.41  \\ [1.0ex]

Flux$_{cptt}$ & 2.06 & 2.02 && 2.14 & 1.98 && 1.67 & 2.05 && 1.83 & 2.08 \\[1.0ex]

Flux$_{total}$ & 2.64 & 2.40 && 2.63 & 2.37 && 2.11 & 2.43 && 2.34 & 2.49   \\  [1.0ex]

$\chi^{2}$/dof & 87/65 & 73/65 && 84/65 & 69/65 && 88/65 & 78/65 && 76/65 & 64/65 \\[1.0ex]

$\nu$ (Hz) & 44.14$^{+1.97}_{-2.00}$ & 35.56$^{+0.33}_{-0.33}$ && 42.15$^{+0.75}_{-0.70}$ & 34.31$^{+0.25}_{-0.26}$ && 27.33$^{+0.19}_{-0.20}$ & 34.71$^{+0.32}_{-0.32}$ && 33.76$^{+0.25}_{-0.27}$ &   37.93$^{+0.38}_{-0.37}$ \\[1.0ex] 

lag (s) & \multicolumn{2}{c}{\text{335$\pm$60}} & & \multicolumn{2}{c}{\text{-59$\pm$40}} & & \multicolumn{2}{c}{\text{702$\pm$80}} & & \multicolumn{2}{c}{\text{-194$\pm$27}} \\[1.0ex]

\hline
\end{tabular}
\end{table*}


\begin{table*}
\centering
\caption{Best-fit parameters of the spectra of different segments using the model  $\tt  tbabs(bbody+diskbb+powerlaw)$. $k$T$_{bb}$ is the black body temperature. N$_{bb}$ is the normalization of the blackbody component. $k$T$_{in}$ is the inner disc temperature and N$_{dbb}$ is the normalization of the diskbb model. $\Gamma$ is the power-law photon index. N$_{pl}$ is the normalization of the power-law model. Flux is in units of 10$^{-08}$ ergs cm$^{-2}$ s$^{-1}$.}
\begin{tabular}{ccccccccccccccc}
\hline \hline
\multicolumn{1}{c}{Parameter} & \multicolumn{2}{c}{\text{Segment A}} && \multicolumn{2}{c}{\text{Segment B}} && \multicolumn{2}{c}{\text{Segment C}} && \multicolumn{2}{c}{\text{Segment D}} \\ [0.5ex]
\cline{2-3}  \cline{5-6} \cline{8-9} \cline{11-12}
\textbf{} & {HBO1} & {HBO2} & & {HBO1} & {HBO2} && {HBO1} & {HBO2} && {HBO1} & {HBO2} \\
\hline \hline

$k$T$_{bb}$ (keV) & 2.94$^{+0.37}_{-0.23}$          & 2.76$^{+0.10}_{-0.08}$      && 2.81$^{+0.12}_{-0.10}$      & 2.80$^{+0.12}_{-0.10}$      && 2.99$^{+0.13}_{-0.11}$       &  2.75$^{+0.10}_{-0.08}$         &&   2.60$^{+0.06}_{-0.06}$       & 2.76$^{+0.11}_{-0.09}$ \\[1.0ex]

N$_{bb}$          & 0.049$^{+0.019}_{-0.015}$     & 0.084$^{+0.011}_{-0.011}$      && 0.077$^{+0.010}_{-0.011}$ & 0.076$^{+0.011}_{-0.011}$ && 0.026$^{+0.037}_{-0.019}$  &  0.085$^{+0.010}_{-0.011}$       &&   0.047$^{+0.035}_{-0.037}$  & 0.084$^{+0.010}_{-0.010}$ \\[1.0ex]

$k$T$_{in}$ (keV) & 2.16$^{+0.10}_{-0.12}$          & 1.93$^{+0.06}_{-0.06}$    && 1.99$^{+0.06}_{-0.06}$      & 2.01$^{+0.07}_{-0.07}$      && 2.46$^{+0.04}_{-0.04}$       & 1.95$^{+0.07}_{-0.06}$         && 2.19$^{+0.08}_{-0.08}$         & 1.95$^{+0.06}_{-0.06}$ \\[1.0ex]

N$_{dbb}$        & 61.16$^{+11.95}_{-8.30}$          & 92.14$^{+11.06}_{-9.68}$     && 90.53$^{+10.54}_{-9.09}$     & 73.55$^{+9.81}_{-8.45}$     && 24.33$^{+1.60}_{-1.57}$      & 86.39$^{+11.02}_{-9.61}$        && 39.62$^{+2.57}_{-3.00}$         & 90.83$^{+11.16}_{-9.70}$  \\[1.0ex]

R$_{in}$ (km) & 9.95$^{+0.92}_{-0.70}$ & 12.21$^{+0.71}_{-0.65}$ && 12.11$^{+0.68}_{-0.62}$ & 10.91$^{+0.70}_{-0.64}$ && 6.27$^{+0.20}_{-0.20}$ & 11.83$^{+0.73}_{-0.67}$ && 8.01$^{+0.25}_{-0.30}$ & 12.13$^{+0.72}_{-0.66}$ \\ [1.0ex]
R$_{eff}$ (km) & 11.79$^{+1.10}_{-0.82}$ & 10.75$^{+0.84}_{-0.78}$ && 14.34$^{+0.81}_{-0.73}$ & 12.93$^{+0.83}_{-0.76}$ && 7.43$^{+0.24}_{-0.24}$ & 14.01$^{+0.86}_{-0.80}$ && 9.49$^{+0.30}_{-0.36}$ & 14.37$^{+0.85}_{-0.78}$ \\[1.0ex]

$\Gamma$         & 3.61$^{+0.16}_{-0.13}$           & 4.84$^{+0.35}_{-0.34}$      && 4.30$^{+0.27}_{-0.26}$      & 4.30$^{+0.24}_{-0.23}$      && 3.18$^{+0.24}_{-0.25}$       & 4.58$^{+0.31}_{-0.29}$        && 3.04$^{+0.29}_{-0.29}$          & 4.61$^{+0.32}_{-0.30}$ \\[1.0ex]  

N$_{pl}$         & 30.20$^{+7.17}_{-5.32}$          & 76.00$^{+43.72}_{-26.87}$   && 46.79$^{+20.10}_{-13.59}$     & 47.72$^{+17.97}_{-12.63}$     && 19.26$^{+2.26}_{-1.97}$       & 59.61$^{+29.46}_{-19.05}$        && 17.53$^{+2.23}_{-1.85}$          & 62.14$^{+31.87}_{-20.38}$  \\[1.0ex]

Flux$_{bb}$     & 0.36      & 0.63       && 0.57    & 0.56     && 0.19     & 0.64      && 0.36     & 0.62 \\[1.0ex]

Flux$_{dbb}$    & 1.79      & 1.58       && 1.79    & 1.54     && 1.28     & 1.57      && 1.23     & 1.64  \\[1.0ex]

Flux$_{pl}$     & 0.48      & 0.18       && 0.25    & 0.26     && 0.62     & 0.21      && 0.73     & 0.21 \\[1.0ex]

Flux$_{total}$  & 2.64      & 2.40       && 2.63    & 2.37     && 2.11     & 2.43      && 2.34     & 2.49  \\ [1.0ex]

$\chi^{2}$/dof  & 88/65     & 73/65      && 83/65   & 68/65    && 87/65    & 77/65     && 76/65    & 64/65 \\[1.0ex]

$\nu$ (Hz) & 44.14$^{+1.97}_{-2.00}$ & 35.56$^{+0.33}_{-0.33}$ && 42.15$^{+0.75}_{-0.70}$ & 34.31$^{+0.25}_{-0.26}$ && 27.33$^{+0.19}_{-0.20}$ & 34.71$^{+0.32}_{-0.32}$ && 33.76$^{+0.25}_{-0.27}$ &   37.93$^{+0.38}_{-0.37}$ \\[1.0ex] 

lag (s) & \multicolumn{2}{c}{\text{335$\pm$60}} & & \multicolumn{2}{c}{\text{-59$\pm$40}} & & \multicolumn{2}{c}{\text{702$\pm$80}} & & \multicolumn{2}{c}{\text{-194$\pm$27}} \\[1.0ex]

\hline 
\hline
\end{tabular}
\end{table*}

\begin{table*}
\centering
\caption{Simultaneous spectral fit results for segments A to D. For more detail see the text.}
\begin{tabular}{ccccccccccccccc}
\hline \hline
           & \multicolumn{2}{c}{\text{Segment A}} && \multicolumn{2}{c}{\text{Segment B}} && \multicolumn{2}{c}{\text{Segment C}} && \multicolumn{2}{c}{\text{Segment D}} \\ [0.5ex]
\hline\hline

          & \multicolumn{2}{c}{\text{$\chi^{2}$/dof = 10126.30/142}} & & \multicolumn{2}{c}{\text{$\chi^{2}$/dof = 15490.52/142}} & & \multicolumn{2}{c}{\text{$\chi^{2}$/dof = 20501.10/142}} & & \multicolumn{2}{c}{\text{$\chi^{2}$/dof = 5642.56/142}} \\[1.0ex]
         
\cline{2-3}  \cline{5-6} \cline{8-9} \cline{11-12}                    
          & \multicolumn{2}{c}{\text{bbody free}} & & \multicolumn{2}{c}{\text{bbody free}} & & \multicolumn{2}{c}{\text{bbody free}} & & \multicolumn{2}{c}{\text{bbody free}} \\[1.0ex]

          & \multicolumn{2}{c}{\text{$\chi^{2}$/dof = 1313.27/140}} & & \multicolumn{2}{c}{\text{$\chi^{2}$/dof = 4933.55/140}} & & \multicolumn{2}{c}{\text{$\chi^{2}$/dof = 1987.19/140}} & & \multicolumn{2}{c}{\text{$\chi^{2}$/dof = 830.67/140}} \\[1.0ex]
          
          & \multicolumn{2}{c}{\text{F-test= 94.33}} & & \multicolumn{2}{c}{\text{F-test= 149.78}} & & \multicolumn{2}{c}{\text{F-test= 652.16}} & & \multicolumn{2}{c}{\text{F-test= 405.49}} \\[1.0ex]
          
         & \multicolumn{2}{c}{\text{Probability= 1.13$\times$10$^{-26}$}} & & \multicolumn{2}{c}{\text{Probability= 1.64$\times$10$^{-35}$}} & & \multicolumn{2}{c}{\text{Probability= 1.13$\times$10$^{-71}$}} & & \multicolumn{2}{c}{\text{Probability= 5.71$\times$10$^{-59}$}} \\[1.0ex] 

\cline{2-3}  \cline{5-6} \cline{8-9} \cline{11-12}                    
          & \multicolumn{2}{c}{\text{diskbb free}} & & \multicolumn{2}{c}{\text{diskbb free}} & & \multicolumn{2}{c}{\text{diskbb free}} & & \multicolumn{2}{c}{\text{diskbb free}} \\[1.0ex]

          & \multicolumn{2}{c}{\text{$\chi^{2}$/dof = 478.88/140}} & & \multicolumn{2}{c}{\text{$\chi^{2}$/dof = 477.80/140}} & & \multicolumn{2}{c}{\text{$\chi^{2}$/dof = 408.87/140}} & & \multicolumn{2}{c}{\text{$\chi^{2}$/dof = 389.44/140}} \\[1.0ex]
          
          & \multicolumn{2}{c}{\text{F-test= 1410.21}} & & \multicolumn{2}{c}{\text{F-test= 2199.44}} & & \multicolumn{2}{c}{\text{F-test= 3439.86}} & & \multicolumn{2}{c}{\text{F-test= 944.22}} \\[1.0ex]
   
          & \multicolumn{2}{c}{\text{Probability= 1.71$\times$10$^{-93}$}} & & \multicolumn{2}{c}{\text{Probability= 1.74$\times$10$^{-106}$}} & & \multicolumn{2}{c}{\text{Probability= 9.69$\times$10$^{-120}$}} & & \multicolumn{2}{c}{\text{Probability= 5.34$\times$10$^{-82}$}} \\[1.0ex]
       
\cline{2-3}  \cline{5-6} \cline{8-9} \cline{11-12}                    
          & \multicolumn{2}{c}{\text{powerlaw free}} & & \multicolumn{2}{c}{\text{powerlaw free}} & & \multicolumn{2}{c}{\text{powerlaw free}} & & \multicolumn{2}{c}{\text{powerlaw free}} \\[1.0ex]

          & \multicolumn{2}{c}{\text{$\chi^{2}$/dof = 763.64/140}} & & \multicolumn{2}{c}{\text{$\chi^{2}$/dof = 2237.90/140}} & & \multicolumn{2}{c}{\text{$\chi^{2}$/dof = 1745.82/140}} & & \multicolumn{2}{c}{\text{$\chi^{2}$/dof = 812.94/140}} \\[1.0ex]
          
          & \multicolumn{2}{c}{\text{F-test= 763.64}} & & \multicolumn{2}{c}{\text{F-test= 414.53}} & & \multicolumn{2}{c}{\text{F-test= 752.00}} & & \multicolumn{2}{c}{\text{F-test= 415.86}} \\[1.0ex]
     
         & \multicolumn{2}{c}{\text{Probability= 2.63$\times$10$^{-79}$}} & & \multicolumn{2}{c}{\text{Probability= 1.52$\times$10$^{-59}$}} & & \multicolumn{2}{c}{\text{Probability= 1.30$\times$10$^{-75}$}} & & \multicolumn{2}{c}{\text{Probability= 1.26$\times$10$^{-59}$}} \\[1.0ex]

\hline
\end{tabular}
\end{table*}


\begin{table*}
\caption {Best-fit parameters of SXT + LAXPC spectra of different parts (p1 to p6) of the HID using the model $\tt tbabs(bbody+compTT)$. Various parameters are the same as discussed in the above-mentioned tables. Flux is in units of 10$^{-08}$ ergs cm$^{-2}$ s$^{-1}$.}
\begin{tabular}{cccccccccccccc}
\hline \hline
Parameter & p1 & p2 & p3 & p4 & p5 & p6 \\
\hline \hline
$k$T$_{bb}$ (keV) & 1.20$^{+0.03}_{-0.03}$ & 1.29$^{+0.04}_{-0.05}$ & 1.31$^{+0.04}_{-0.04}$ & 1.28$^{+0.05}_{-0.05}$ & 1.30$^{+0.06}_{-0.06}$ & 1.24$^{+0.09}_{-0.08}$ \\[1.0ex]
N$_{bb}$ & 0.100$^{+0.004}_{-0.004}$ & 0.099$^{+0.006}_{-0.005}$ & 0.103$^{+0.007}_{-0.006}$ & 0.086$^{+0.006}_{-0.005}$ & 0.077$^{+0.006}_{-0.005}$ & 0.046$^{+0.005}_{-0.005}$  \\[1.0ex]
kT$_{0}$ (keV) & 0.26$^{+0.04}_{-0.05}$ & 0.26$^{+0.04}_{-0.04}$ & 0.26$^{+0.04}_{-0.04}$ & 0.25$^{+0.04}_{-0.05}$ & 0.28$^{+0.05}_{-0.04}$ & 0.31$^{+0.06}_{-0.07}$ \\[1.0ex]
$k$T$_e$ (keV) & 2.58$^{+0.10}_{-0.08}$ & 2.73$^{+0.09}_{-0.08}$ & 2.73$^{+0.10}_{-0.08}$ & 2.73$^{+0.09}_{-0.08}$ & 2.76$^{+0.10}_{-0.08}$ & 2.76$^{+0.11}_{-0.08}$ \\[1.0ex]
$\tau$ & 6.46$^{+0.36}_{-0.39}$ & 7.17$^{+0.37}_{-0.35}$ & 7.18$^{+0.40}_{-0.41}$ & 7.29$^{+0.38}_{-0.38}$ & 7.23$^{+0.40}_{-0.43}$ & 6.91$^{+0.44}_{-0.47}$ \\[1.0ex]
N$_{cptt}$ & 1.34$^{+0.29}_{-0.14}$ & 1.45$^{+0.22}_{-0.15}$ & 1.38$^{+0.17}_{-0.13}$ & 1.44$^{+0.28}_{-0.16}$ & 1.26$^{+0.15}_{-0.13}$ & 1.18$^{+0.24}_{-0.11}$  \\[1.0ex]
Flux$_{bb}$ & 2.03 &  2.06 & 2.11 &  1.76 & 1.59 & 1.02 \\[1.0ex]
Flux$_{cptt}$ & 2.11 & 2.98 & 2.83 & 2.98 & 2.85 & 2.85 \\[1.0ex]
Flux$_{total}$ & 4.14 & 5.04 & 4.94 & 4.74 & 4.44 & 3.87 \\[1.0ex]
$\chi^{2}$/dof & 504/496 & 436/496 & 491/496 & 457/496 & 499/496 & 477/476 \\[1.0ex]
\hline
\end{tabular}
\end{table*}

\begin{table*}
\caption{Same as above table except that the model used to unfold the spectra is $\tt tbabs(diskbb+compTT)$.  Flux is in units of 10$^{-8}$ ergs cm$^{-2}$ s$^{-1}$.  Luminosity is in units of 10$^{38}$ ergs s$^{-1}$ assuming distance 9 kpc and the inclination angle is 60$^{\circ}$ of GX 5-1. Mass accretion rate (\.{m}) is in units of 10$^{18}$ g s$^{-1}$.}
\begin{tabular}{cccccccccccccc}
\hline \hline
Parameter & p1 & p2 & p3 & p4 & p5 & p6 \\
\hline \hline
$k$T$_{in}$ (keV) & 1.16$^{+0.02}_{-0.02}$ & 1.18$^{+0.03}_{-0.03}$ & 1.23$^{+0.03}_{-0.03}$ & 1.33$^{+0.03}_{-0.03}$ & 1.22$^{+0.03}_{-0.03}$ & 1.27$^{+0.03}_{-0.03}$ \\[1.0ex]
N$_{dbb}$ & 231.31$^{+22.06}_{-21.96}$ & 222.76$^{+22.07}_{-21.87}$ & 189.55$^{+20.54}_{-18.21}$ & 154.93$^{+13.14}_{-13.69}$ & 183.54$^{+19.72}_{-19.69}$ &  157.04$^{+17.89}_{-16.00}$ \\[1.0ex]
R$_{in}$ (km) & 19.35$^{+0.90}_{-0.94}$ & 18.99$^{+0.91}_{-0.95}$ & 17.52$^{+0.92}_{-0.86}$ & 15.84$^{+0.65}_{-0.71}$ & 17.24$^{+0.90}_{-0.95}$ & 15.95$^{+0.88}_{-0.83}$ \\ [1.0ex]
R$_{eff}$ (km) & 22.93$^{+1.06}_{-1.11}$ & 22.50$^{+1.08}_{-1.13}$ & 20.76$^{+1.09}_{-1.02}$ & 18.77$^{+0.77}_{-0.84}$ & 20.43$^{+1.06}_{-1.12}$ & 18.89$^{+1.04}_{-0.98}$ \\[1.0ex]
kT$_{0}$(keV) & 1.25$^{+0.04}_{-0.04}$ & 1.32$^{+0.05}_{-0.06}$ & 1.33$^{+0.05}_{-0.05}$ & 1.36$^{+0.08}_{-0.07}$ & 1.34$^{+0.07}_{-0.06}$ &  1.37$^{+0.13}_{-0.13}$ \\[1.0ex]
$k$T$_e$ (keV) & 3.17$^{+0.47}_{-0.26}$ & 3.09$^{+0.27}_{-0.18}$ & 3.06$^{+0.26}_{-0.18}$ & 2.99$^{+0.27}_{-0.15}$ & 3.03$^{+0.26}_{-0.16}$ & 2.87$^{+0.27}_{-0.15}$ \\[1.0ex]
$\tau$ & 7.19$^{+1.03}_{-1.29}$ & 9.01$^{+1.05}_{-1.17}$ & 9.08$^{+1.10}_{-1.18}$ & 10.37$^{+1.26}_{-1.58}$ & 10.11$^{+1.19}_{-1.43}$ &  12.43$^{+2.15}_{-2.32}$\\[1.0ex]
N$_{cptt}$ & 0.63$^{+0.07}_{-0.09}$ & 0.78$^{+0.05}_{-0.07}$ & 0.78$^{+0.05}_{-0.07}$ & 0.66$^{+0.04}_{-0.06}$ & 0.67$^{+0.04}_{-0.05}$ &  0.45$^{+0.02}_{-0.04}$ \\[1.0ex]
Flux$_{dbb}$ & 1.61 & 1.65 & 1.67 & 1.90 & 1.55 & 1.71  \\[1.0ex]
Flux$_{cptt}$ & 2.53 & 3.40 & 3.27 & 2.84 & 2.90 & 2.16  \\[1.0ex]
Flux$_{total}$ &  4.14 & 5.05 & 4.94 & 4.74 & 4.45 & 3.87\\[1.0ex]
L$_{0.8-20}$ & 4.01 & 4.89 & 4.78 & 4.59 & 4.31 & 3.75 \\ [1.0ex]
\.{m} & 2.15 & 2.63 & 2.57 & 2.47 & 2.32 & 2.01 \\[1.0ex]
$\chi^{2}$/dof & 522/496 & 458/496 & 518/496 & 475/496 & 513/496 & 483/476 \\[1.0ex]
\hline
\end{tabular}
\end{table*}

\begin{table*}
\centering
\caption{Same as above table except that the model used to unfold the spectra is  $\tt tbabs(bbody+diskbb+powerlaw)$. Flux is in units of 10$^{-08}$ ergs cm$^{-2}$ s$^{-1}$.}
\begin{tabular}{ccccccccccccccc}
\hline \hline

Parameter & p1 & p2 & p3 & p4 & p5 & p6 \\

\hline \hline

$k$T$_{bb}$ (keV) & 1.12$^{+0.03}_{-0.03}$          & 1.21$^{+0.05}_{-0.04}$      & 1.24$^{+0.04}_{-0.04}$      & 1.20$^{+0.05}_{-0.06}$      & 1.20$^{+0.06}_{-0.06}$       &  1.05$^{+0.10}_{-0.11}$         \\[1.0ex]

N$_{bb}$          & 0.079$^{+0.009}_{-0.011}$     & 0.072$^{+0.015}_{-0.008}$      & 0.083$^{+0.008}_{-0.011}$ & 0.064$^{+0.009}_{-0.013}$  & 0.054$^{+0.010}_{-0.012}$  &  0.034$^{+0.008}_{-0.009}$      \\[1.0ex]

$k$T$_{in}$ (keV) & 2.47$^{+0.12}_{-0.12}$          & 2.84$^{+0.16}_{-0.09}$    & 2.93$^{+0.07}_{-0.13}$      & 2.92$^{+0.10}_{-0.15}$      & 2.91$^{+0.12}_{-0.15}$       & 2.89$^{+0.13}_{-0.15}$       \\[1.0ex]

N$_{dbb}$        & 11.35$^{+4.21}_{-2.95}$          & 9.67$^{+1.90}_{-2.47}$     & 8.07$^{+2.12}_{-1.16}$     & 8.53$^{+2.92}_{-1.50}$     & 8.35$^{+2.75}_{-1.65}$      & 7.21$^{+2.96}_{-1.39}$        \\[1.0ex]

R$_{in}$ (km) & 4.28$^{+0.73}_{-0.59}$ & 3.95$^{+0.37}_{-0.54}$ & 3.61$^{+0.44}_{-0.26}$ & 3.71$^{+0.58}_{-0.34}$ & 3.67$^{+0.56}_{-0.38}$ & 3.41$^{+0.61}_{-0.34}$  \\ [1.0ex]

R$_{eff}$ (km) & 5.08$^{+0.86}_{-0.70}$ & 4.68$^{+0.44}_{-0.64}$ & 4.28$^{+0.52}_{-0.31}$ & 4.40$^{+0.69}_{-0.40}$ & 4.35$^{+0.66}_{-0.45}$ & 4.04$^{+0.75}_{-0.41}$ \\[1.0ex]

$\Gamma$         & 2.77$^{+0.21}_{-0.16}$           & 2.64$^{+0.48}_{-0.17}$      & 2.85$^{+0.33}_{-0.29}$      & 2.71$^{+0.43}_{-0.29}$      & 2.60$^{+0.38}_{-0.27}$       & 2.68$^{+0.43}_{-0.27}$        \\[1.0ex]  

N$_{pl}$         & 1.48$^{+0.38}_{-0.36}$          & 1.31$^{+0.46}_{-0.27}$   & 1.48$^{+0.31}_{-0.38}$     & 1.33$^{+0.34}_{-0.38}$     & 1.13$^{+0.32}_{-0.35}$       & 1.27$^{+0.43}_{-0.42}$       \\[1.0ex]

Flux$_{bb}$     & 1.61      & 1.49       & 1.70    & 1.31     & 1.10     & 0.72       \\[1.0ex]

Flux$_{dbb}$    & 2.01      & 3.01       & 2.78    & 2.93     & 2.85     & 2.63       \\[1.0ex]

Flux$_{pl}$     & 0.55      & 0.58       & 0.51    & 0.54     & 0.51     & 0.57       \\[1.0ex]

Flux$_{total}$  & 4.20      & 5.09       & 4.99    & 4.78     & 4.48     & 3.96       \\ [1.0ex]

$\chi^{2}$/dof  & 508/496     & 446/496      & 505/496   & 463/496    & 510/496    & 486/476      \\[1.0ex]

\hline
\end{tabular}
\end{table*}

\clearpage
  \section*{ACKNOWLEDGEMENTS}     
  
We thank the referee for the comments and suggestions that have improved the quality of the paper. CP and KS acknowledge the Indian Space Research Organization (ISRO) for providing financial support under the AstroSat archival data utilization program.  This work uses the data from the AstroSat mission of the Indian Space Research Organization (ISRO), archived at the Indian Space Science Data Center (ISSDC). This work has used the data from the Soft X-ray Telescope (SXT) developed at TIFR, Mumbai, and the SXT POC at TIFR is thanked for verifying and releasing the data via the ISSDC data archive and providing the necessary software tools.  This work has also used the data from the LAXPC Instruments developed at TIFR, Mumbai, and the LAXPC-POC at TIFR is thanked for verifying and releasing the data. We also thank the AstroSat Science Support Cell hosted by IUCAA and TIFR for providing the LAXPCSOFT software which we used for LAXPC data analysis.

  \section*{DATA AVAILABILITY}
  
Data underlying this article are available at AstroSat-ISSDC website
\url{(https://astrobrowse.issdc.gov.in/astro_archive/archive/Home.jsp)}

\bibliographystyle{mnras}
\bibliography{references}

\bsp
\label{lastpage}
\end{document}